\documentclass[showpacs,pre,aps,twocolumn,superscriptaddress]{revtex4-1}

\pdfoutput=1

\usepackage{amsmath,amsfonts,amssymb,bm,graphicx,hyperref}
\usepackage{color}

\newcommand{\eps}{\varepsilon}

\begin{document}

\title{Overlap and activity glass transitions in plaquette spin models with hierarchical dynamics}

\author{Robert M. Turner}
\affiliation{School of Physics and Astronomy, University of Nottingham, Nottingham, NG7 2RD, United Kingdom}

\author{Robert L. Jack}
\affiliation{Department of Physics, University of Bath, Bath, BA2 7AY, United Kingdom}

\author{Juan P. Garrahan}
\affiliation{School of Physics and Astronomy, University of Nottingham, Nottingham, NG7 2RD, United Kingdom}

\date{\today}

\begin{abstract}
We consider thermodynamic and dynamic phase transitions
in plaquette spin models of glasses.  
The thermodynamic transitions involve coupled (annealed) replicas of the model.  We map these coupled-replica systems to a single replica in a magnetic field,
which allows us to analyse the resulting phase transitions in detail.  For the triangular plaquette model (TPM), 
we find for the coupled-replica system a phase transition between high- and low-overlap phases, occuring at a 
coupling $\eps^*(T)$, which vanishes in the low-temperature limit.
Using computational path sampling techniques,
we show that a single TPM also displays ``space-time''  transitions between active and inactive dynamical phases.  These first-order dynamical transitions occur at a critical counting field $s_{c}(T) \gtrsim 0$ that appears to vanish at zero temperature, in a manner reminiscent of the thermodynamic overlap transition.  In order to extend the ideas to three dimensions we introduce the square pyramid model which also displays both overlap and activity transitions. We discuss a possible common origin of these various phase transitions, based on long-lived (metastable) glassy states.
\end{abstract}

\maketitle

\section{Introduction}

As supercooled liquids approach their glass transitions, one observes a very sharp increase in their viscosities and structural relaxation times.  
The physical mechanism
underlying this slow dynamics remains controversial~\cite{Ediger1996,Cavagna2009,Berthier2011,Biroli2013}.  Some theories, particularly
the random first-order transition theory~\cite{Lubchenko2007}, propose
that glassy systems are approaching some kind of thermodynamic phase transition, with associated collective (slow) dynamics.  The existence
of such phase transitions can be probed by computing the free-energy of a pair of coupled copies (or replicas) of the system, and searching for a transition 
as a function of both temperature and coupling strength.
These transitions link an equilibrium-like phase where the replicas are different from each other (the liquid) to one where they become very 
similar (the glass)~\cite{Franz1997}.  The similarity bewteen the configurations is measured by an overlap variable, which is the order parameter
for the transition.
An alternative approach, that of dynamical facilitation~\cite{Chandler2010}, links glassy behaviour to a dynamical ``space-time'' phase transition.  
This transition is explored through distributions of time-integrated observables, which quantify activity in the dynamics~\cite{Garrahan2007,Hedges2009}.  Based on these distributions, one
may infer the existence of transitions between an active dynamical phase (the equilibrium liquid) and an inactive phase (the non-equilibrium glass).  The order parameter
for these transitions is the dynamical activity.

In this work, we investigate plaquette spin models of glasses~\cite{Garrahan2002}, 
for which both overlap-fluctuations and dynamical activity-fluctuations can be analysed, by a combination of analytical and
computational methods.  We concentrate on two models, whose relaxation behaviour is similar to that of the facilitated East model~\cite{Jackle1991,Ritort2003,Garrahan2003} --
their relaxation times increase faster than an Arrhenius law at low temperatures, but the equilibrium relaxation time is finite at all 
positive temperatures, diverging only as $T\to0$.
We present evidence that these models support \emph{both} dynamic and thermodynamic phase transitions.  In the thermodynamic case we consider a coupling between two \emph{annealed} replicas, and transitions {\em occur only for non-zero (positive) values of the coupling}.  

We argue that these results provide a connection between the (apparently quite different) `thermodynamic' and `dynamic' theories of the glass transition. This connection
is built on the idea of metastability, which is intrinsically connected to glassy behaviour. The formation of a metastable state in a finite
dimensional system requires that small perturbations in that state do not grow: the system prefers to relax back into the metastable state.
This stability to small perturbations may be described in terms of an interfacial cost that acts to penalise local perturbations.
Different theories ascribe different origins to these interfacial costs, which might be either static
or dynamic, depending on the system of interest, and the kinds of fluctuation being considered.  However, the existence of these interfacial
costs seems quite generic, and may be useful for rationalising different kinds of phase transition in these systems.

The main results of this work are as follows.  We analyse the triangular plaquette model (TPM) in two spatial dimensions~\cite{Newman1999,Garrahan2000}, and
a three-dimensional variant of this model, which we refer to as the square pyramid model (SPyM).  
In Section~\ref{sec:theory}, we show that two (annealed) coupled replicas of these systems can be mapped to a single replica in a magnetic field, and we derive some useful features of this single-replica system, which place constraints on the kind of phase transitions that can occur.  Some of these results were derived in previous work~\cite{Heringa1989,Sasa2010,Garrahan2014}, but our analysis contains several new insights.  In Section~\ref{sec:numeric-field}, we show numerical
evidence that the TPM in a magnetic field supports a phase transition in the $2d$-Ising universality class. It then follows from the mappings in Section~\ref{sec:theory} that the
coupled replicas of these systems also support a similar phase transition.  In Section~\ref{sec:s-ens-tpm}, we show that the TPM also supports dynamical
``space-time'' phase transitions, similar to those in~\cite{Garrahan2007,Hedges2009,Elmatad2010}.  In Section~\ref{sec:spym}, we introduce the SPyM, and show evidence that it supports phase transitions
in the coupled replica setting, and dynamical space-time phase transitions.  Finally in Section~\ref{sec:discuss}, we discuss the relationships between the
thermodynamic and dynamical phase transitions that we have found, and we consider the consequences of these results for theories of the glass transition.

\section{Plaquette models, coupled replicas, and mapping to system in field}
\label{sec:theory}

\subsection{Models}

We consider plaquette spin models defined in terms of classical Ising spins on regular lattices, with energy functions of the form  
\begin{equation}
E_J(\sigma) \equiv -\frac{J}{2} \sum_{\mu} \sigma_{i_{\mu}} \sigma_{j_{\mu}} \cdots \sigma_{k_{\mu}} ,
\label{E}
\end{equation}
where $\sigma_{i}=\pm 1$ with $i$ indicating a lattice site ($i = 1,\ldots,N$), and where the interactions are in terms of products of spins $\sigma_{i_{\mu}} \sigma_{j_{\mu}} \cdots \sigma_{k_{\mu}}$ around the {\em plaquettes} $\mu$ of the lattice.  
See~\cite{Newman1999,Garrahan2000,Garrahan2002} for a more general overview of the relevant properties
of these systems.
On a square lattice, one labels each square
plaquette with an index $\mu$, and $\{\sigma_{i_{\mu}}, \sigma_{j_{\mu}}, \ldots, \sigma_{k_{\mu}}\}$ is the set of four spins on the vertices of plaquette $\mu$.
This construction is easily generalised to higher dimensions: for a cubic lattice and cubic ``plaquettes'', each term in the energy would involve eight spins.
This motivates us to define plaquette variables $\tau_\mu = \sigma_{i_{\mu}} \sigma_{j_{\mu}} \cdots \sigma_{k_{\mu}}$.

An interesting model in this class is the TPM~\cite{Garrahan2002},  where the lattice is triangular and the interactions are between triplets of spins in the corners of upward pointing triangles, 
\begin{equation}
E_J(\sigma) \equiv -\frac{J}{2} \sum_{\mu = \vartriangle} \sigma_{i_{\mu}} \sigma_{j_{\mu}} \sigma_{k_{\mu}} , \;\;\; (\text{TPM})
\label{ETPM}
\end{equation}
Our analysis rests on a correspondence between configurations of the spin variables $\sigma_i$ and the plaquette variables $\tau_\mu$.  If we first consider
rhombus-shaped systems whose linear size is an integer power of $2$, with periodic boundaries, then there is a one-to-one mapping between 
spin configurations and plaquette configurations.  (It is clear that every spin configuration corresponds to a unique plaquette configuration, but the 
existence of a spin configuration corresponding to \emph{every} plaquette configuration is less trivial \cite{Newman1999,Garrahan2000,Garrahan2002,Ritort2003}.)  
For systems of different sizes or with different boundary conditions, the correspondence is not perfectly one-to-one, but these deviations turn out to be 
irrelevant in the thermodynamic limit.
In Section~\ref{sec:spym} below, we will also discuss the SPyM, a three-dimensional model with the same one-to-one correspondence, on the body-centred cubic (bcc) lattice.

In cases where the one-to-one mapping holds exactly,
the fully polarised state $\sigma_i=1 \; \forall i$ is the unique ground state of (\ref{E}).
In terms of the plaquette variables, the ground state is $\tau_{\mu} = 1 \; \forall \mu$, and the elementary excitation is a ``defect'', $\tau_{\mu} = -1$. 
Two-body spin correlations vanish in these models~\cite{Garrahan2002}, although higher-order
spin correlations are finite and allow access to a growing length scale at low temperatures~\cite{Jack2005caging}.
Also, since there is a one-to-one mapping between spins and plaquettes,
the thermodynamic properties of these models are those non-interacting binary plaquette variables \cite{Newman1999,Garrahan2000,Garrahan2002,Ritort2003},
or a free gas of `defective' plaquettes (with $\tau_\mu=-1$)
\cite{[{If the plaquette interactions do not allow for a spin-defect duality, as for example with spins in a cubic lattice with interactions on the square faces (rather than on the cubes), then the static properties may be non-trivial. See for example }] [{, and references therein.}] Johnston2012}.

However, while the thermodynamic properties of plaquette models are trivial, their (single spin-flip) dynamics is not. This effect arises because flipping a single spin $\sigma_i$
changes the states of all of the plaquettes in which it participates.  The plaquette dynamics is therefore ``kinetically constrained'' \cite{Garrahan2000,Garrahan2002,Ritort2003,[{This is even true in mean-field versions of plaquette models, see }]Foini2012}, possibly leading  to complex glassy dynamics at low temperatures.  This is what occurs for example in the TPM whose dynamical properties are similar to those of the East facilitated model \cite{Garrahan2000,Garrahan2002,Ritort2003}, displaying ``parabolic'' super-Arrhenius relaxation, dynamic heterogeneity, and other characteristic features of the glass transition \cite{Biroli2013}.

\subsection{Coupled replicas}

To probe thermodynamic overlap fluctuations, we consider
two coupled replicas of a plaquette model~\cite{Franz1997,Franz2013,Berthier2013,Parisi2013,Garrahan2014}. The energy function of the combined system is
\begin{equation}
E_{J,\varepsilon}(\sigma^{a},\sigma^{b}) \equiv E_J(\sigma^{a}) + E_J(\sigma^{b}) - \varepsilon \sum_{i} \sigma_{i}^{a} \sigma_{i}^{b},
\label{E2}
\end{equation}
where $\sigma^{a}$ and $\sigma^{b}$ are the spin configurations in the replicas $a$ and $b$. The overlap,
\begin{equation}
%q(\sigma^{a},\sigma^{b}) \equiv N^{-1} \sum_{i} \sigma_{i}^{a} \sigma_{i}^{b} ,
Q(\sigma^{a},\sigma^{b}) \equiv \sum_{i} \sigma_{i}^{a} \sigma_{i}^{b} ,
\label{q}
\end{equation}
measures how similar the two copies are, and the strength of their coupling given by its conjugate field $\varepsilon$.
The coupling (\ref{E2}) is denoted {\em annealed} since both replicas are allowed to fluctuate on an equal footing.  The case of {\em quenched} coupling, in contrast, involves one of the replicas being frozen in an equilibrium configuration.  Here we will only consider the case of annealed coupling which is easier to treat both analytically and numerically.  
Hence the partition function for these two coupled replicas is
\begin{equation}
%Z_2(J,\varepsilon) = \sum_{\{\sigma^a,\sigma^b\}} e^{-\beta E_{J,\varepsilon}(\sigma^{a},\sigma^{b})} .
Z_2(J,\varepsilon) = \sum_{\sigma^a,\sigma^b} e^{-\beta E_{J,\varepsilon}(\sigma^{a},\sigma^{b})} .
\label{Z2}
\end{equation}
where the sum runs over the configurations $\sigma^a,\sigma^b$: that is, over all $\sigma^a_i=\pm1$
and all $\sigma^b_i=\pm1$.  Here and in the following, we sometimes set $\beta=1$ where there is no ambiguity [for example, the left hand side of (\ref{Z2})
should strictly be $Z_2(\beta J,\beta\varepsilon)$ but we suppress the dependence on $\beta$, for simplicity].

\begin{figure}[t!]
\includegraphics[width=0.8\columnwidth]{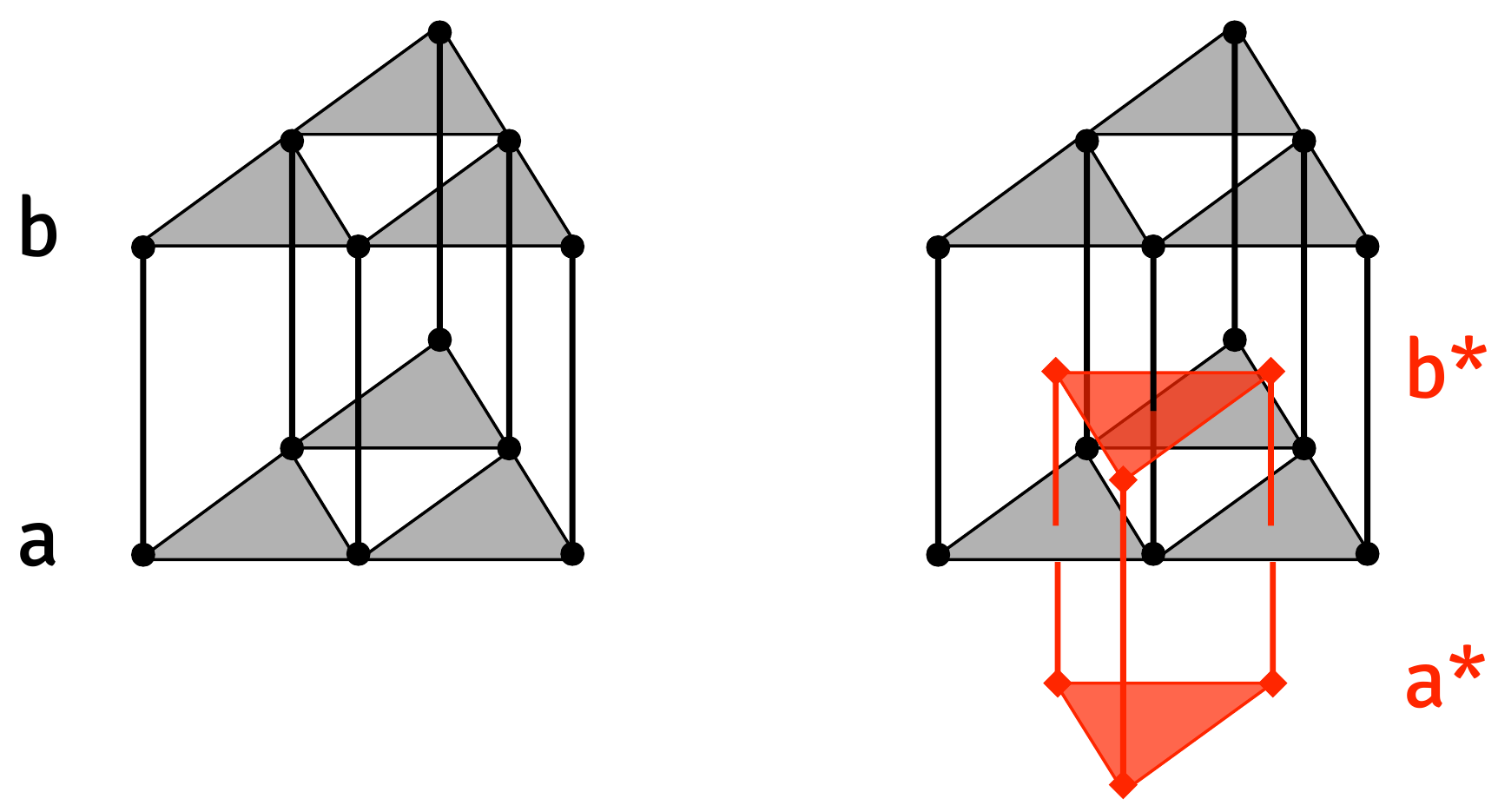}
\caption{(Color online) Sketch of the duality relation in the TPM.  The left panel shows two TPM systems, $a$ and $b$, with coupling as in \eqref{E2}.  The right panel shows the location of the sites of the dual problem, again two coupled TPMs $a*$ and $b*$.  The plaquettes in the dual system bisect the coupling interactions in the direct system, and vice-versa, giving rise to the duality transformation \eqref{Z2d2}.
}
\label{fig-tpm-layer}
\end{figure}

\subsection{Mapping to single system in field}

In~\cite{Garrahan2014}, a mapping was derived between the free energies of the coupled system (\ref{Z2}) and a single plaquette model
in a magnetic field.  Here, we present a mapping between (sets of) configurations of these systems, which extends that analysis,
as well as recovering the same mapping between free energies.

We introduce overlap variables $q_i=\sigma_i^a \sigma_i^b$ on each site: our aim is to calculate the statistical
weight of a particular configuration of these variables.  This weight is
\begin{equation}
W_2(q|J,\varepsilon) =  \sum_{\sigma^a,\sigma^b} e^{-\beta E_{J,\varepsilon}(\sigma^{a},\sigma^{b})} \prod_i \delta{(q_i -\sigma^a_i\sigma^b_i)} .
\label{wZ2q}
\end{equation}
%We introduce overlap variables $q_i$ on each site, and write
%\begin{equation}
%Z_2(J,\varepsilon) =  \sum_{\{q\}} \sum_{\{\sigma^a,\sigma^b\}} e^{-\beta E_{J,\varepsilon}(\sigma^{a},\sigma^{b})} \prod_i \delta{(q_i -\sigma^a_i\sigma^b_i)} .
%\label{Z2q}
%\end{equation}
We now perform the sum over the $\sigma$ variables.  If we sum over $\sigma^b$ first we obtain,
\begin{multline}
W_2(q|J,\varepsilon) =  \sum_{\sigma^a} 
\exp\left[\frac{\beta J}{2}\sum_\mu 
\sigma^a_{i_{\mu}} \sigma^a_{j_{\mu}} \cdots \sigma^a_{k_{\mu}} 
\right.
\nonumber \\
\left.
\times
\left( 1 + q_{i_{\mu}} q_{j_{\mu}} \cdots q_{k_{\mu}} \right)
 + \beta \varepsilon \sum_i q_i
 \right] .
\end{multline}
%\begin{eqnarray}
%Z_2(J,\varepsilon) =  \sum_{\{q\}} \sum_{\{\sigma^a\}} 
%\exp\left[\frac{\beta J}{2}\sum_\mu 
%\sigma^a_{i_{\mu}} \sigma^a_{j_{\mu}} \cdots \sigma^a_{k_{\mu}} 
%\right.
%\nonumber \\
%\left.
%\times
%\left( 1 + q_{i_{\mu}} q_{j_{\mu}} \cdots q_{k_{\mu}} \right)
% + \beta \varepsilon \sum_i q_i
% \right] .
%\end{eqnarray}
For the summation over $\sigma^a$ we  replace $\sigma^a_{i_{\mu}} \sigma^a_{j_{\mu}} \cdots \sigma^a_{k_{\mu}}$ by $\tau^a_\mu$.
Then we use the characteristic feature of the model, that plaquette and spin configurations are in a one-to-one correspondence, so 
we replace the sum over the $\sigma^a_i$ with a sum over the $\tau^a_\mu$. 
\begin{multline}
W_2(q|J,\varepsilon) = \sum_{\tau^a} 
\exp\left[\frac{\beta J}{2}\sum_\mu 
\tau^a_\mu
\left( 1 + q_{i_{\mu}} q_{j_{\mu}} \cdots q_{k_{\mu}} \right)
\right.
\nonumber \\
\left.
 + \beta \varepsilon \sum_i q_i
 \right] ,
 \nonumber
\end{multline}
%\begin{eqnarray}
%Z_2(J,\varepsilon) =  \sum_{\{q\}} \sum_{\{\tau^a\}} 
%\exp\left[\frac{\beta J}{2}\sum_\mu 
%\tau^a_\mu
%\left( 1 + q_{i_{\mu}} q_{j_{\mu}} \cdots q_{k_{\mu}} \right)
%\right.
%\nonumber \\
%\left.
% + \beta \varepsilon \sum_i q_i
% \right] ,
% \nonumber
%\end{eqnarray}
Performing the sum, we arrive at
\begin{equation}
W_2(q|J,\varepsilon) =  (4\cosh\beta J)^{N/2}  \cdot
e^{-\beta E_{J'}(q) + \beta \varepsilon \sum_i q_i} .
\label{wZ2q2}
\end{equation}
%\begin{equation}
%Z_2(J,\varepsilon) =  2^N \cosh^{N/2}(\beta J)  
%\sum_{\{q\}}
%e^{\beta E_{J'}(q) + \beta \varepsilon \sum_i q_i} .
%\label{Z2q2}
%\end{equation}
with 
\begin{equation}
\beta J' = \log \cosh(\beta J) 
\label{JpJ}
\end{equation}
We recognise the exponential term in (\ref{wZ2q2}) as the statistical weight of a 
configuration $\sigma=q$ for a single plaquette model with energy scale $J'$, in a magnetic field $h=\epsilon$.

To explore the consequences of this property for the free energy,
we observe $Z_2(J,\eps) = \sum_{q} W_2(q|J,\eps)$,  so that
%
%
%This means that the partition sum of two replicas, with energy constant $J$ and coupling $\varepsilon$, is proportional to the partition sum of a single plaquette system with energy constant $J'$ and magnetic field $h$ where,
%\begin{equation}
%\beta J' = \log \cosh(\beta J) 
%\, , \,\,\,
%h = \varepsilon .
%\label{Jp}
%\end{equation}
%That is,
\begin{equation}
Z_2(J,\varepsilon) =  (4\cosh\beta J)^{N/2}\cdot Z_{1}(J',\varepsilon) ,
\label{Z21}
\end{equation}
where 
\begin{equation}
Z_1(J,h) =    
%\sum_{\{\sigma\}}
\sum_{\sigma}
e^{-\beta E_{J}(\sigma) + \beta h \sum_i \sigma_i} .
\end{equation}
is the partition function of a single plaquette model in a field $h$.
In addition, this latter system is known to have an exact duality \cite{Heringa1989,Sasa2010}
\begin{equation}
Z_1(J,h) = (\sinh\beta J \sinh2\beta h)^{N/2} Z_1(\tilde{J},\tilde{h}) ,
\label{Z1d1}
\end{equation}
where
\begin{equation}
e^{-\beta \tilde{J}} = \tanh(\beta h) \, , \,\,\,
e^{-2 \beta \tilde{h}} = \tanh(\beta J/2) .
\label{Zhd2}
\end{equation}
From Eqs.\ \eqref{Z21}-\eqref{Zhd2} the duality of the coupled plaquette system follows:
\begin{equation}
Z_2(J,\varepsilon) = (\sinh\beta J \sinh \beta \varepsilon)^N Z_2(J^*,\varepsilon^*) ,
\label{Z2d1}
\end{equation}
with
\begin{equation}
 e^{-\beta \varepsilon^*} = \tanh(\beta J/2)    \, , \,\,\,
%e^{-\beta \varepsilon^*} = \tanh(\beta J/2) .
 e^{-\beta J^*} = \tanh(\beta \varepsilon/2) .
\label{Z2d2}
\end{equation}
This duality is precisely the one obtained in \cite{Garrahan2014} for the two coupled replicas of the TPM.  
(Note that if $\tanh y = e^{-2x}$ then $\tanh x = e^{-2y}$, which follows from the definition of the tanh function,
and facilitates inversion of these duality transforms.)

\subsection{Dualities and phase transitions}

The mapping from two coupled replicas to a single system in a field has useful consequences, since we may exploit existing
results for plaquette models in magnetic fields.   
The duality of this model, Eq.\ \eqref{Z1d1} (see also~\cite{Sasa2010}), implies a duality relation for the free energy $F_1=-\log Z_1$,
\begin{equation}
F_1(h,J) + \frac{N}{2} \log \sinh(2h) = F_1(\tilde{h},\tilde{J}) + \frac{N}{2} \log \sinh(2\tilde{h})
\label{equ:dual}
\end{equation}
where $\tilde{h}$ and $\tilde{J}$ are given in Eq.\ \eqref{Zhd2}, and we set $\beta=1$ as above, without loss of generality.
Phase transitions appear as singularities in the free energy density $f_1=\lim_{N\to\infty} F_1/N$.
For any given $J$, it follows that if there is a single phase-transition at some $h$, it must happen on the self-dual line $(h,J)=(\tilde{h},\tilde{J})$, for which 
\begin{equation}
J = -\log \tanh h .
\label{self-dual-field}
\end{equation} 
On this line one has also 
$\sinh J \sinh 2h=1$.  
Such phase transitions were investigated
in~\cite{Heringa1989,Sasa2010}: the plaquette models considered there support a single critical point that occurs at some point $(J_c,h_c)$ on this line,
with first-order phase coexistence occurring on the part of the line with $J>J_c$.

From the above mapping, these transitions correspond to phase transitions in the coupled replica system: the first-order transition line
separates a state with low overlap (small $\eps$) from one with high overlap (large $\eps$).  For the coupled replicas, the self-dual line is
\begin{equation}
%\sinh(\beta J) \sinh(\beta h) = 1 ,
\sinh(\beta J) \sinh(\beta \eps) = 1 ,
\label{selfdual}
\end{equation}
This situation, where the self-dual line for the coupled-replica system
contains a first-order transition region and a critical point, was proposed for the TPM in Ref.\ \cite{Garrahan2014}.
We present numerical evidence for this situation in Section~\ref{sec:numeric-field} below.

\subsection{Other consequences of dualities and symmetries}

In this section, we explore some further consequences of the results derived thus far.
First, 
we note that the relation (\ref{wZ2q2}) means that for a coupled
replica system at parameters $(J,\eps)$, the probability of a particular configuration of the overlap variables $q$ is the same as the
probability of finding the configuration $\sigma=q$ for a single system in a field, with parameters $(J',h=\eps)$.  From a numerical perspective,
the single system in a field is much simpler to simulate, and the result (\ref{wZ2q2}) means that such a simulation
provides direct access to all observables based on the overlap variables.  
(This result is much stronger than a mapping at the level of free energies.)

Second, for a geometical interpretation of the duality relation (\ref{Z2d2}), we refer to Figure~\ref{fig-tpm-layer}.  The original coupled system can be thought of as a lattice consisting of two parallel layers, $a$ and $b$. 
The duality relation (\ref{Z2d2}) may be interpreted as a mapping between two different two-layer systems, where the plaquette
energy scale in one model determines the interlayer coupling in the other, and vice versa.  Fig.~\ref{fig-tpm-layer}(b) illustrates
this situation, in which the interlayer `bonds' in the original system intersect the intralayer plaquettes in the dual
system, and \emph{vice versa}.
 This geometrical way of seeing the duality easily generalises to other lattices and plaquette interactions.

\begin{figure}
\includegraphics[width=\columnwidth]{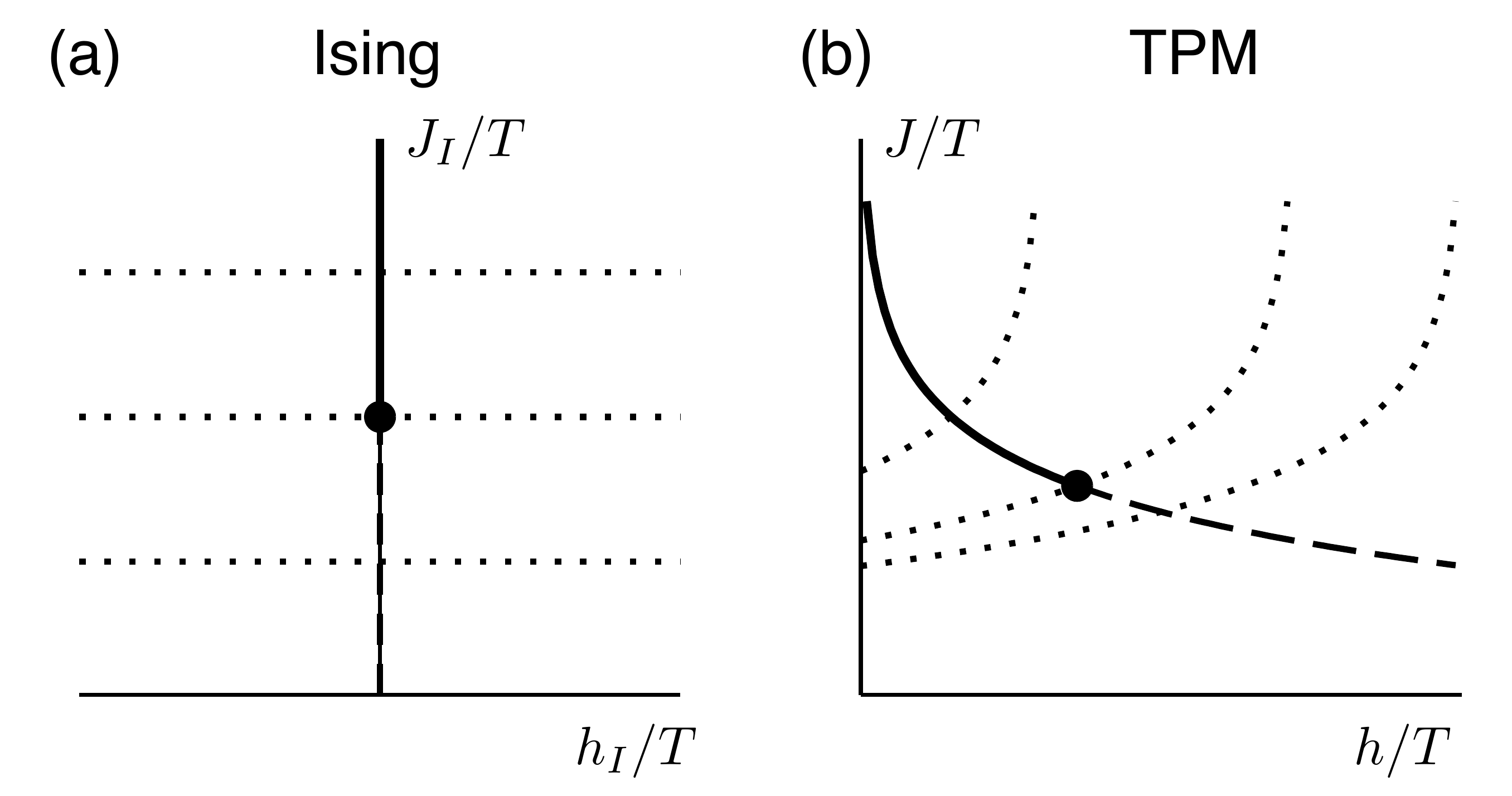}
\caption{Illustration of the relation between critical behaviour of the Ising model and the TPM.  (a)~Ising phase diagram. 
the $h_I=0$ axis is a symmetry line, there is a critical point indicated by a circle, with a first-order (phase coexistence) line
for large $J_I$, indicated by a solid line.  Selected lines of constant $J_I$ are indicated by dotted lines.  
(b)~The corresponding situation for the TPM in a field.  On the solid/dashed line (\ref{self-dual-field}), the system has a discrete ($Z_2$) symmetry: 
a critical point and phase coexistence both occur on this line, as indicated.  The dotted lines are obtained from (\ref{eq:Jh0}) for three
different values of $h_0$, and correspond to the lines of constant $J_I$ in panel (a).  Near the critical point, they indicate the direction
of the most relevant renormalisation group flow.}
\label{fig-ising-tpm}
\end{figure}

\begin{figure*}[t!]
\includegraphics[width=1.9\columnwidth]{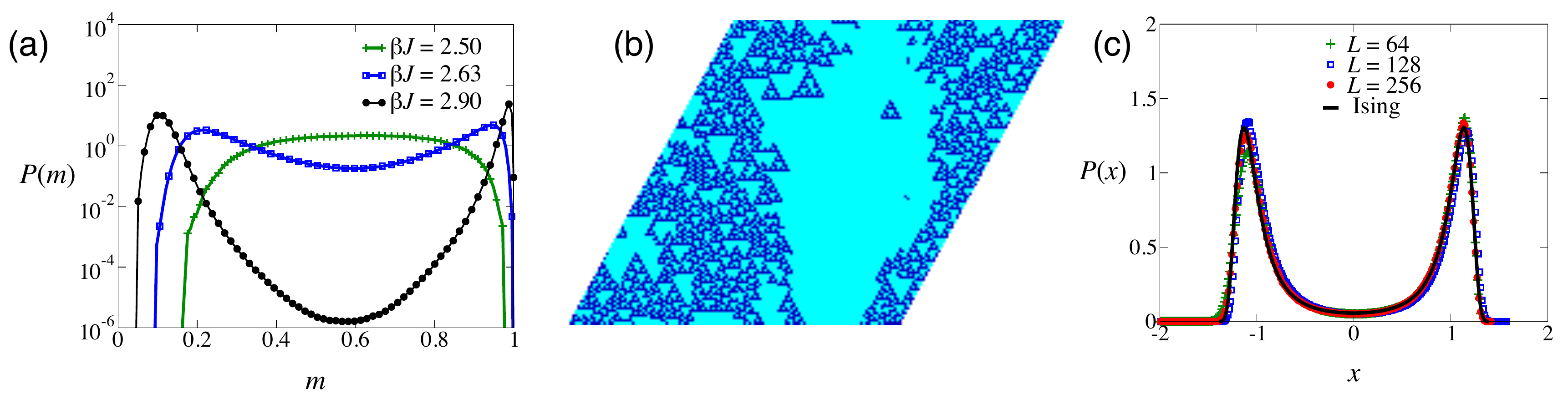}
\caption{(Color online)
Simulations of the TPM in a field. (a) Distribution of the magnetisation
at various values of $J$ for state points on the self-dual line \eqref{selfdual}, at system size $L=128$.  The bimodal distribution $P(m)$ indicates a first-order transition, which disappears on reducing $J$.  From (\ref{wZ2q2}), the same distributions would be obtained when considering the overlap between two coupled TPMs, at appropriate state points. (b) Representative configuration at phase coexistence ($\beta J=2.9$ and $L=128$) showing interfaces between regions of small and large magnetisation (corresponding to regions of small and large overlap in the two-replica problem).  (c) At our estimated critical point, $(J_c=2.634,\, h_c=0.072)$ and for various system sizes, we show distributions of the variable $x$ that is obtained by rescaling the order parameter $\cal M$ to zero mean and unit variance.  The full line is the corresponding result for the $2d$ Ising model at criticality~\cite{Nico1988}, indicating that the critical point of the TPM in a field (and therefore of the two coupled TPMs) is in the $2d$ Ising universality class. }
\label{fig-tpm-field}
\end{figure*}

Third, the duality relation for a plaquette model in a field can be used to analyse the behaviour of its free
energy in the vicinity of a (presumed) critical point.
Given (\ref{equ:dual}), it is useful to define the ``singular part'' of the free energy of a plaquette model in a field
\begin{equation}
F_{\rm sing}(h,J) = F_1(h,J) + \frac{N}{2} \log \sinh(2h)
\end{equation}
We assume that a critical point exists somewhere on the self-dual line, and 
that this critical point is in the Ising universality class, as is found generically~\cite{Heringa1989}
(see also below).
We can then consider a renormalisation group flow near the critical point, which will have
two relevant directions in the $(h,J)$ plane.     
One of these relevant directions corresponds to the ferromagnetic coupling $J_I$ of an associated Ising model; the other to 
a magnetic field $h_I$ in the Ising model.

For the plaquette model, the relation (\ref{equ:dual}) is associated with a symmetry that will be spontaneously broken
at the phase transition.  Hence, the relevant direction that corresponds to the Ising coupling 
must be the one that preserves the self-dual symmetry: this direction
lies along the self-dual line.  However, it will be useful in the following to also identify the direction that corresponds to the Ising field $h_I$.
To accomplish this, we define a family of curves in the $(h,J)$ plane that correspond to lines of constant $J_I$ in the Ising system. The family of
curves is parameterised by the points $(h_0,J_0)$ where they cross the self-dual line.  The curves are invariant under the duality transformation
(just as lines of constant $J_I$ are invariant under the Ising symmetry transformation $h\to-h$.)  Furthermore, the singular part of the free energy, 
calculated along each curve,
is an even function of $h-h_0$ (just as the Ising free energy is even in $h_I$, for constant $J_I$).

The relevant family of curves is given by the formula
\begin{equation}
J_{h_0}(h) = -\log \tanh( 2h_0 - h) .
\label{eq:Jh0}
\end{equation}
The dual of any point on a line $J_{h_0}(h)$ is easily verified to be $(\tilde{J},\tilde{h}) = ( J_{h_0}(2h_0-h) , 2 h_0 - h )$, which also satisfies (\ref{eq:Jh0}), confirming
that the line is invariant.  It then follows from (\ref{equ:dual}) that $F_{\rm sing}$ is a symmetric function of $h-h_0$.
The resulting situation is shown in Fig.~\ref{fig-ising-tpm}.

The significance of the curves (\ref{eq:Jh0}) is that they give the direction of the most relevant RG flow at the critical point.  The most natural
order parameter for the phase transition is then obtained by taking a derivative of the free energy along these lines, for which
\begin{equation}
\frac{\mathrm{d}}{\mathrm{d}h} = \frac{\partial}{\partial h} + \frac{\partial J_{h_0}}{\partial h} \partial_J = \frac{\partial}{\partial h} + 2 \sinh J \frac{\partial}{\partial J} .
\end{equation}
We therefore define the order parameter
\begin{equation}
{\cal M} = -\frac{\mathrm{d}}{\mathrm{d}h} [F+ (NJ/2)] = M - 2N_d\sinh J ,
\label{M}
\end{equation}
where $M=-(\partial F/\partial h)=\sum_i \sigma_i$ is the magnetisation, and 
$N_d=(\partial/\partial J)[F+(NJ/2)]=\frac12 \sum_\mu (1-\tau_\mu)$ is the number of defective plaquettes.
A key advantage of this order parameter is that only even cumulants of $\cal M$ may
show singular behaviour at phase transitions: the odd cumulants can be shown to depend only 
on derivatives of the non-singular $\log \sinh 2h$ term in the free energy.  The use of this order
parameter also accounts automatically for ``field-mixing''~\cite{Wilding1995} when analysing
critical properties.

\section{Numerical results for the TPM in a field}
\label{sec:numeric-field}

We performed numerical simulations of the TPM in a field, to analyse its phase behaviour.  Working always on the self-dual line (\ref{selfdual}) we use continuous time Monte Carlo simulations~\cite{Bortz1975,NewmanBook}
to sample a reweighted Boltzmann distribution $P(\sigma) \propto b(M(\sigma)) e^{-\beta E(\sigma)}$, where $b(M)$ is a bias function.  We measure the resulting distribution $P_{\rm b}(M)$ of the magnetisation, but we choose the function $b(M)$ so that this sampled distribution does not include any deep minima (free energy barriers)~\cite{bruce2003}.  The `true' distribution $P(M)$ associated with the unbiased model is then easily obtained as $P(M) \propto P_{\rm b}(M) / b(M)$.

Figure~\ref{fig-tpm-field}(a) shows numerical evidence that for large $J$ (and small $h$), the self-dual line is associated with a first-order phase transition: see also Ref.\ \cite{Sasa2010}. The figure shows the distribution $P(m)$ of the magnetisation density $m=M/N$, for three state points on the self-dual line.  At large $J$ the distribution is bimodal, characteristic of first-order coexistence.  A typical configuration at these conditions is shown in Fig.\ 2(b), showing coexistence of low- and high-$m$ regions, separated by sharp interfaces, as expected for a first-order transition.  Based on smaller systems, a previous study~\cite{Sasa2010} speculated that phase separation would not occur for the TPM in a field, but our results show that this does indeed occur  large enough systems are considered.
Given the mapping (\ref{wZ2q2}), Fig.~\ref{fig-tpm-field}(b) is also a representative configuration of the overlap between two
coupled replicas, for suitable $(J,\eps)$.

\begin{figure}
\includegraphics[width=\columnwidth]{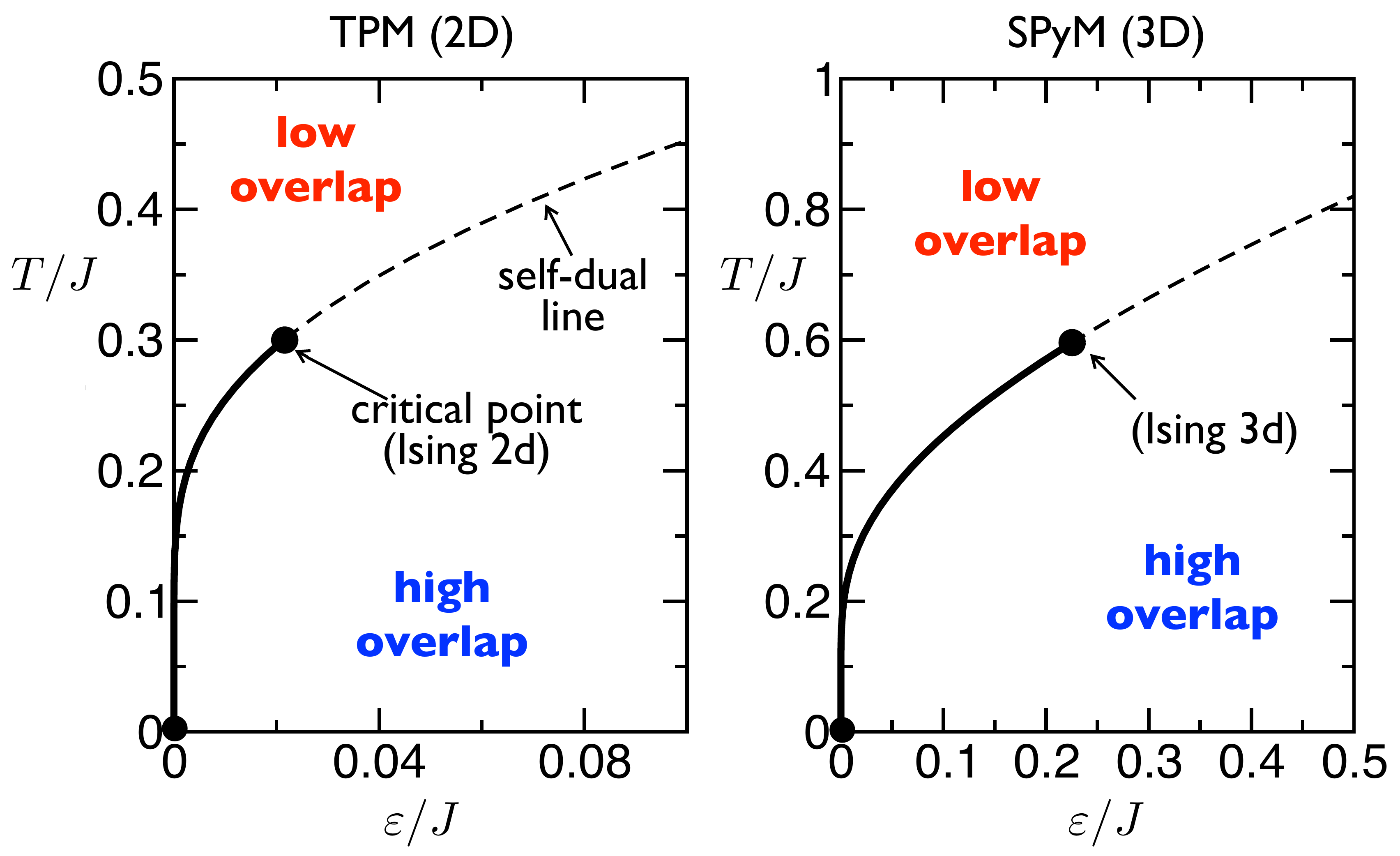}
\caption{(Color online) Phase diagrams of coupled plaquette models, the two-dimensional TPM (left) and the three-dimensional SPyM (right).  The full line corresponds to a line of first-order transitions between a thermodynamic phase of small overlap and one of large overlap between the replicas.  This curve is on the self-dual line \eqref{selfdual} (dashed line).  The first-order transition line ends at a critical point that is in the $2d$ Ising universality class for the TPM and the $3d$ Ising universality class for the SPyM.}
\label{fig-eps-phase}
\end{figure}

As $J$ is decreased (or equivalently, temperature and field are increased) along the self-dual line, the bimodality in $P(M)$ becomes less pronounced and eventually disappears.  
Fig.~\ref{fig-tpm-field}(a) indicates that the first-order line terminates at a critical point 
$(J_c,h_c)$ with $J_c\approx 2.6$.
To identify the universality class of this phase transition, we performed a finite-size scaling analysis, using the order parameter
$\cal M$ defined in (\ref{M}). 
Note that if all spins are up, one has $M=N$ and $N_d=0$, giving ${\cal M}=N$.  On the other hand, in a
state with $h=0$ then $M=0$ and $N_d=1/(1+e^J)$, at low temperatures this gives ${\cal M}\approx-N$.  In general, one expects a crossover between these two limits as $h$ is increased from $0$, with the crossover occuring near ${\cal M}=0$.
Figure 2(c) shows the distribution of the order parameter $\cal M$ at our numerically estimated critical point, $(J_c=2.634,\, h_c=0.072)$, for various system sizes $L$.  
These distributions are rescaled to have a mean of zero and a variance of unity: the variable 
$x=({\cal M} - \langle {\cal M} \rangle)/\sqrt{\langle({\cal M} - \langle {\cal M} \rangle)^2 \rangle}$.  The key point is that 
the shapes of the curves do not depend on the system size, as expected at a critical point.  

These critical distributions
are universal: Fig.~\ref{fig-tpm-field}(c) also shows the corresponding data for a two-dimensional Ising model at the critical
point~\cite{Nico1988}.  The distributions match well, indicating that the system is in the Ising univerality class.  
We also calculated the ratio of the susceptibilities $\chi=L^{-d} \langle\delta{\cal M}^2\rangle$ at criticality, for the system sizes $L=128$ and $L=256$.  We find
$\chi({L=256})/\chi({L=128})=3.36$. Theory predicts that this ratio should scale as $L^{2Y_H-d}$ where $Y_H$ is the
exponent associated with the field conjugate to the order parameter~\cite{Heringa1989}.  For $2d$-Ising universality, one has $Y_H=15/8$ 
which yields a prediction of $2^{7/4}\approx 3.364$ for the ratio of susceptibilities.  Hence, these results are also consistent with $2d$-Ising universality.

Bringing together these results, we arrive at the phase diagram shown in Fig.~\ref{fig-eps-phase}.  We re-introduce the temperature $T=\beta^{-1}$ as an explicit parameter and plot the phase diagrams as a function
of $T/J$ and $\eps/J$ since this is conventional representation in supercooled (glassy) liquids.  The form of this phase diagram
was proposed in~\cite{Garrahan2014}.  However, the results here are not consistent with the arguments presented in that work for the location and universality class of the critical point.  Our finding that the critical point is in the  Ising class is interesting, since this is the expected result from other general arguments \cite{Franz2013}, and what is observed in simulation of coupled liquids \cite{Parisi2013,Berthier2013}.  However, a key feature of the plaquette models is that the first-order transition line does
not intersect the $\eps=0$ axis except at $T=0$~\cite{Garrahan2014}.

\section{Active-inactive dynamical transitions in the TPM}
\label{sec:s-ens-tpm}

The TPM falls into a category of glassy models that are thermodynamically simple but where glassy behaviour arises because of non-trivial dynamical pathways to the equilibrium state at low temperature \cite{Fredrickson1984,Jackle1991,Garrahan2000,Ritort2003}.  In fact, the dynamics of the TPM \cite{Garrahan2000} is closely related to that of a two-dimensional East model \cite{Jackle1991,Ritort2003,Garrahan2003}.  Many kinetically constrained models, including the East model, display dynamical phase transitions -- phase transitions in the space of trajectories -- between a phase with a high dynamical activity, $K$, and one with low dynamical activity \cite{Garrahan2007}.  
For these lattice models, the activity $K$ is defined as the total number of configuration changes in a trajectory \cite{Lecomte2007,Baiesi2009}.
By coupling a field $s$ to the dynamical activity one can define the so-called $s$-ensemble (also known as the exponentially biased or tilted ensemble) where the probability of obtaining a trajectory $X_{\tau}$ of total time extension $\tau$, is re-weighted by its activity \cite{Lecomte2007,Garrahan2007,Hedges2009} 
\begin{equation}
\label{s-ens}
P_s(X_{\tau}) = \frac{e^{-sK} P_0(X_{\tau})}{Z_s(\tau)} .
\end{equation}
Here $P_s(X_{\tau})$ is the probability of the trajectory $X_\tau$ in the $s$-ensemble, $P_0(X_{\tau})$ is the unbiased probability of this trajectory (the one generated by the actual dynamics of the system), and $Z_s(\tau)$ is the moment generating function of the activity $K$, in effect a partition sum for trajectories. For long times $Z_s(\tau)$ takes on a large-deviation form \cite{Lecomte2007,Garrahan2007,Touchette2009},
\begin{equation}
\label{MGF}
Z_s(\tau) \propto e^{-\tau \theta(s)},
\end{equation}
where $\theta(s)$ plays the role of %(minus) 
a dynamical free-energy, and is the scaled cumulant generating function of the activity. The mean time-intensive activity of trajectories in the $s$-ensemble, 
\begin{equation}
k(s) \equiv \langle K \rangle_{s}/\tau,
\label{equ:ks}
\end{equation}
 where the average is w.r.t.\ the ensemble \eqref{s-ens}, can thus be obtained from $k(s) = \frac{d}{ds}\theta(s)$.  Dynamical -- or ``space-time'' -- phase transitions manifest as singularities in $\theta(s)$ \cite{Lecomte2007,Garrahan2007,Hedges2009,Speck2012}.  We also define the susceptibilty
 \begin{equation}
 \chi(s) \equiv -\frac{\mathrm{d}k}{\mathrm{d}s} = \tau^{-1} \langle \delta K^2 \rangle_s .
 \end{equation}

Like the East model, the dynamical relaxation of the TPM is hierarchical, due to energy barriers to relaxation that are logarithmic in the linear size of relaxing regions \cite{Ritort2003}.  This in turn leads to a ``parabolic'' \cite{Elmatad2009} super-Arrhenius law for the typical relaxation time in equilibrium as a function of temperature. Given the similar dynamical properties of the TPM and the East model dynamics, a natural question is whether the TPM also displays active-inactive ``space-time'' transitions.  

In order to answer this question numerically we make use of transition path sampling (TPS) \cite{Bolhuis2002} to efficiently sample trajectories in the $s$-ensemble. For the purposes of numerical efficiency we exploit the fact that different trajectory ensembles can be defined by fixing different dynamical quantities, such that for long trajectories these distinct ensembles become equivalent \cite{Chetrite2013,Budini2014}.  This is analogous to what occurs with standard configuration ensembles in equilibrium statistical mechanics in the large size limit.  We consider in particular the $x$-ensemble introduced in \cite{Budini2014}, that is, the ensemble of trajectories with fixed number of configuration changes, i.e.\ activity, $K$ but where the overall trajectory time extension $\tau$ fluctuates.  In contrast, the $s$-ensemble above is one where trajectories are of fixed overall time but their activity fluctuates.  For large $K$ and $\tau$ these ensembles can be shown to be equivalent \cite{Budini2014,Kiukas2015}.  For the case of the TPM the $x$-ensemble is particularly efficient to simulate (see \cite{Budini2014} and \cite{Turner2014} for details) and the functions $k(s)$ and $\chi(s)$ can be recovered from the ensemble equivalence. 

\begin{figure}[t!]
\includegraphics[width=0.95\columnwidth]{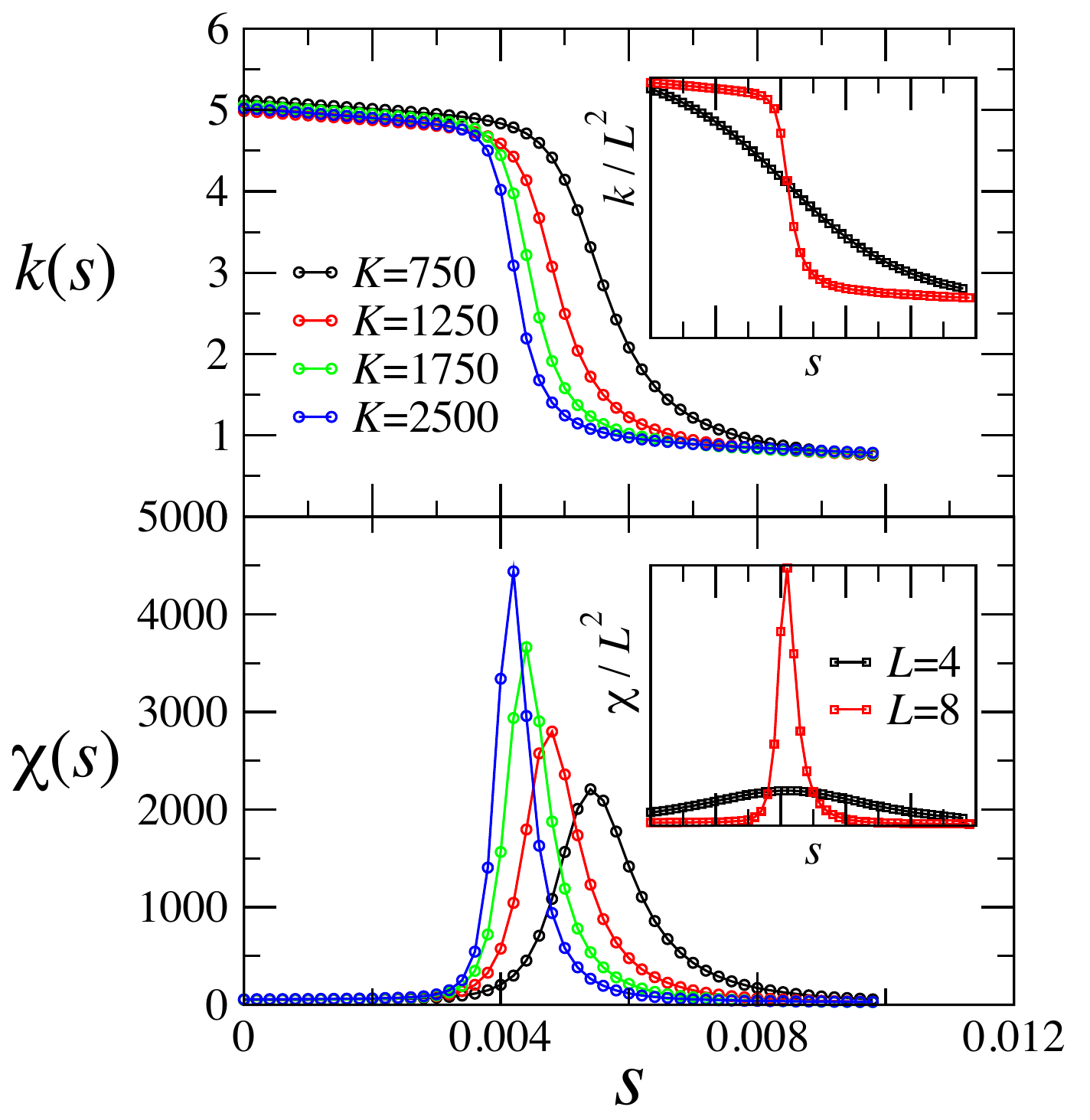}
\caption{(Color online) Average activity $k(s)$, and the associated susceptibility $\chi(s)$ in the TPM at $T=0.5$.  
The main panels show data for system size $L=8$.   These results were obtained by the $x$-ensemble method (see text),
using trajectories with fixed numbers of events $K$, as shown.  On increasing the trajectory length,
the crossover from active to inactive behaviour becomes increasingly sharp and the susceptibility peak increases.
The behaviour for smaller systems ($L=4$) is shown
in the insets, with both quantities normalised by the system size. 
In the absence of a phase transition, one expects 
both $k(s)/L^d$ and $\chi(s)/L^d$ to be independent of $L$, so the 
sharper crossover at $L=8$ is again consistent with an underlying phase transition.}
\label{fig-tpm-s-ens}
\end{figure}

Figure~\ref{fig-tpm-s-ens} shows the results of the $s$-ensemble analysis of the TPM.  It shows the average activity for a system of size $N=8 \times 8$ as a function of $s$, at temperature $T=0.5$ (note this is the TPM in the absence of field).  As the length of the trajectories is increased the change in $\langle K \rangle_{s}$ becomes more pronounced, as seen in the corresponding susceptibilities.  This is indicative of a first-order transition at some $s_{c} \gtrsim 0$.  Similar size scaling is observed by changing system size, as shown in the insets.  The dependence of this transition on the temperature is discussed in Sec.~\ref{sec:spym-e-ens} below, together with similar results for the (three-dimensional) SPyM.

\section{Overlap and activity transitions in a three-dimensional plaquette model}
\label{sec:spym}

In order to explore whether the static and dynamical transitions found above for the TPM are present in dimensions other than two it is of interest to generalise the TPM to higher dimensions.  One of the reasons is that if one wishes to consider plaquette models to study
``quenched'' coupled replicas~\cite{Franz1997,Franz2013,Berthier2013} or ``random pinning''~\cite{Cammarota2012,Jack2012,Kob2013} the distinction between two and three dimensions may be very significant, due to the inability of two-dimensional systems in random fields to support
first-order static transitions \cite{Aizenman1989}.  Here we introduce a three-dimensional model that is similar to the TPM.

\subsection{Model}

\begin{figure}
\includegraphics[width=.5\columnwidth]{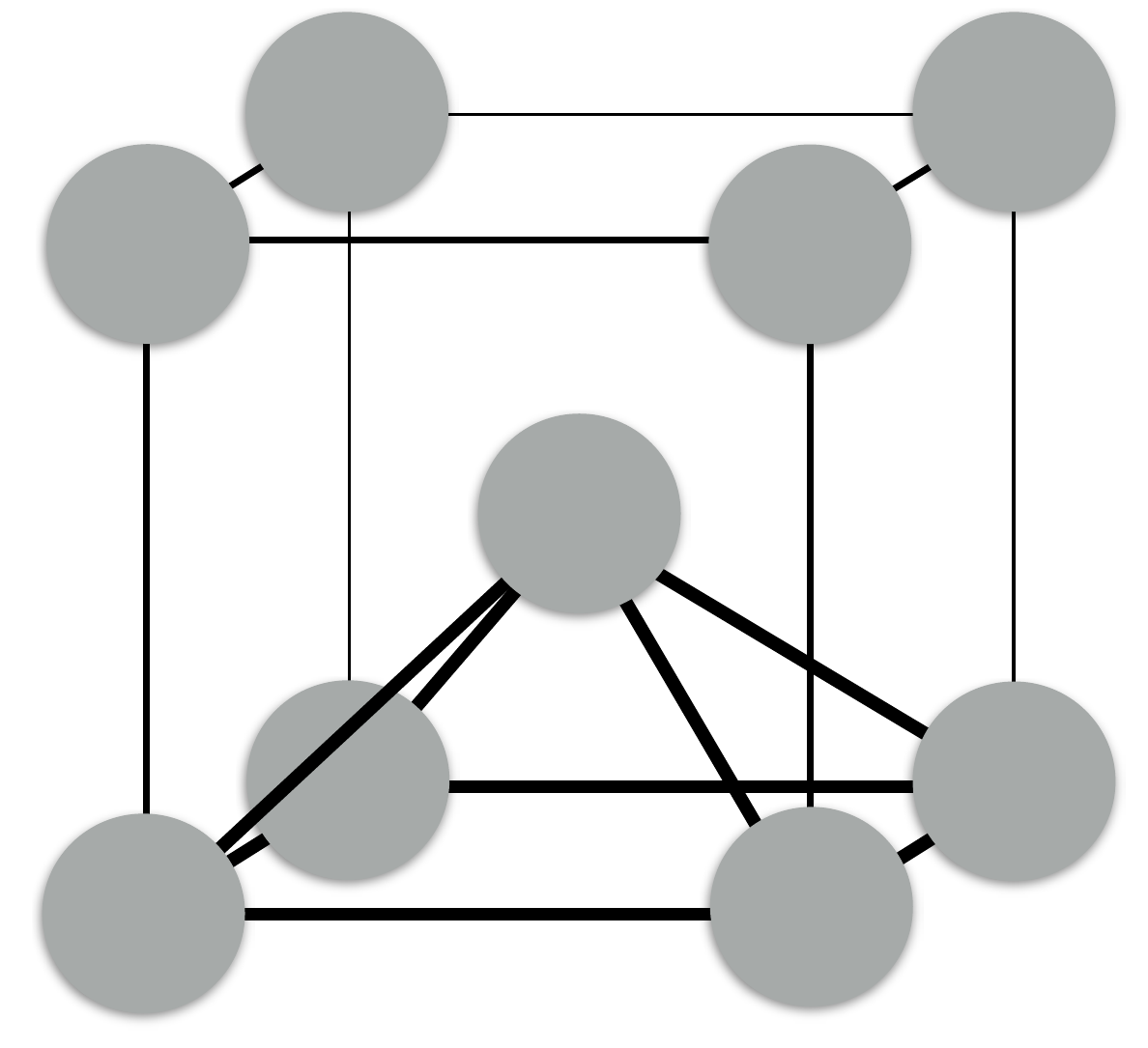}
\caption{(a) The SPyM consists of spins (grey circles) on the sites of a BCC lattice, which interact 
in quintuplets at the vertices of upward pointing square pyramids.  One such pyramid
is indicated; the central spin also participates in four other upward pointing pyramids whose apexes are the four spins on the upper face of the cube.%
}
\label{fig-bcc}
\end{figure}

The model we consider is defined on a three-dimensional body-centred cubic (bcc) lattice.  The ``plaquettes'' are upward-pointing square-based pyramids, each containing five spins.  
Considering the standard bcc unit cell, one such pyramid is formed by the spin at the 
centre of the cube together with the four spins at the corners of the lower face: see Fig.~\ref{fig-bcc}.
We call this model the square pyramid plaquette model, or SPyM. Its energy function reads,
\begin{equation}
E_J(\sigma) \equiv -\frac{J}{2} \sum_{\mu} %\sigma_{{\mu_{1}}} \sigma_{{\mu_{2}}} \sigma_{{\mu_{3}}} \sigma_{{\mu_{4}}} \sigma_{{\mu_{5}}} , 
\sigma_{i_\mu} \sigma_{j_\mu} \sigma_{k_\mu} \sigma_{l_\mu} \sigma_{m_\mu}
\;\;\; (\text{SPyM})
\label{ETPM}
\end{equation}
where $\mu$ runs over all the pyramidal plaquettes on the lattice, and the location of the five interacting spins $\sigma_{i_\mu} \cdots \sigma_{m_\mu}$ is shown in Fig.~\ref{fig-bcc}.  This is ``model 1'' of~\cite{Heringa1989}.

Just like the TPM, the SPyM has a one-to-one correspondence between spin and plaquette configurations.  
An alternative model~\cite{Heringa1989,Garrahan2002} may be defined on a face-centred cubic lattice, in which the plaquettes are tetrahedral pyramids (``model 2'' of~\cite{Heringa1989}).  However, in this case each
interaction involves four spins so the system has a global spin-flip symmetry, and the spin-plaquette correspondence is not exact in finite (periodic) systems.
However, these deviations from the one-to-one correspondence are irrelevant in the thermodynamic limit.

Returning to the SPyM, we explicitly demonstrate the one-to-one correspondence between spins and plaquettes, 
by a general method that applies also to the TPM.  We choose as basis vectors for the lattice $\vec{a}_{1}=(1,0,0)$, $\vec{a}_{2}=(0,1,0)$ and $\vec{a}_{3}=(-1,-1,\sqrt{2})/2$.
We focus on systems whose sites are at $l\vec{a}_1+m\vec{a}_2-n\vec{a}_3$ with $l,m,n \in \{0,1,2,\dots,L-1\}$, with periodic boundaries (so for example
sites with $n=L-1$ are neighbours of those with $n=0$).
We indicate the location of the $\mu$-th pyramid by the position of the spin at the apex. 
The plaquette variable $\tau_{\mu}$ for $\mu=(i,j,k)$ is then
\begin{eqnarray}
\tau_{(i,j,k)} = \sigma_{(i,j,k)}\sigma_{(i,j,k-1)}\sigma_{(i-1,j,k-1)}\sigma_{(i,j-1,k-1)}
\nonumber \\
\times
\sigma_{(i-1,j-1,k-1)} .
\end{eqnarray}
Following the same reasoning as in \cite{Garrahan2002} we can invert this relation in terms of a ``Pascal pyramid'': the idea is to demonstrate that introducing a single defect into the system corresponds to flipping a particular set of spins.  
%Starting from the ground state and repeatedly applying this procedure allows a spin configuration to be generated for every plaquette configuration, which is sufficient for the one-to-one correspondence. 

Starting from the ground state, we demonstrate the procedure by introducing a single defect at the origin:
this affects those spins in upper layers which lie on the sites of an inverted Pascal pyramid (or fractal pyramid). 
Assuming that the central spin in Fig.~\ref{fig-bcc} is at the origin, we flip that spin, which introduces a defect in the pyramid below it.  In
order to avoid any other defects, we also flip the four spins on the top face of the cube shown in Fig.~\ref{fig-bcc}, which ensures that there are no defects
in any of the pyramids pointing upward from the origin.  Iterating this procedure for all other layers, the final spin configuration is
\begin{equation}
\sigma_{(i,j,k)} = 1 - 2 \left[ \left( \begin{array}{c} k \\ i \end{array} \right) \left( \begin{array}{c} k \\ j \end{array} \right) \mod{2} \right] , 
\label{equ:pascal}
\end{equation}
where $\left( \begin{smallmatrix} n\\ r \end{smallmatrix} \right)=\frac{n!}{r!(n-r)!}$ are combinatorial numbers, and $0 \leq i,j \leq k$ (all
other spins have $\sigma=1$). 
Given periodic boundary conditions, this procedure determines all spins in the system: on setting the final layer of spins, it may be
that defects in the final layer are unavoidable.  However, for systems whose linear size $L$ is a power of 2, it is easily shown that
this procedure produces a final state with exactly one defect.
Now observe that for any spin configuration,
flipping the set of spins for which $\sigma=-1$ in (\ref{equ:pascal}) inverts the state of the plaquette variable just below the origin,
leaving all other plaquette variables constant.
Similarly, to flip the state of any other plaquette,
one applies a spatial translation to set of spins for which $\sigma=-1$ in (\ref{equ:pascal}), and flips all the spins within this translated set.
Hence, by repeatedly applying this procedure, one can generate a spin configuration that corresponds to any given configuration of the plaquette variables.
This is sufficient for a one-to-one correspondence between spin and plaquette configurations, since 
we know already that the total number of spin and plaquette configurations are the same, and that every spin 
configuration corresponds to exactly one plaquette configuration.

\begin{figure}[t!]
\includegraphics[width=.9\columnwidth]{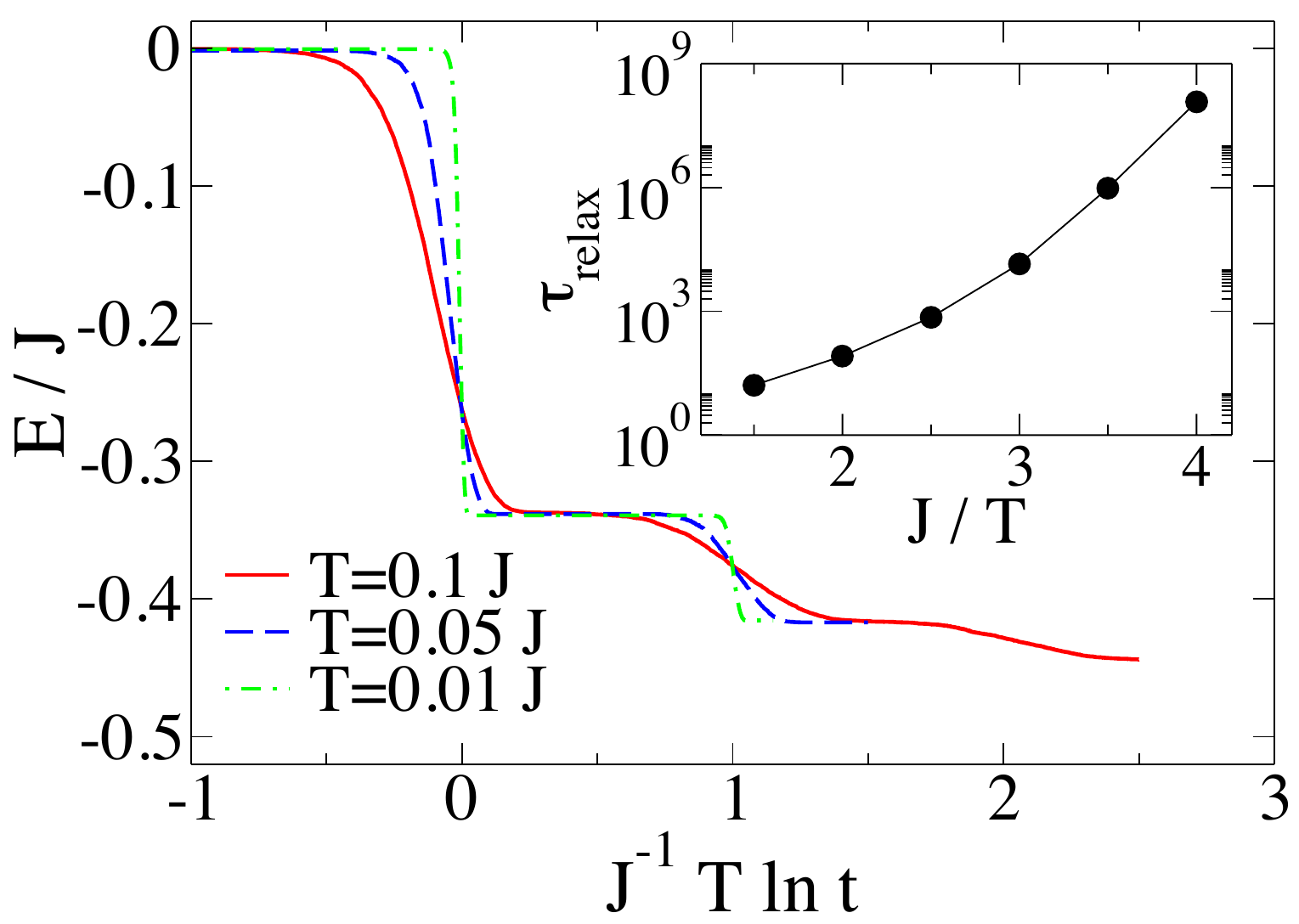}
\caption{(Color online)
Relaxation of the energy of the SPyM at low temperature starting from a random configuration (the system size is $L=16$).  
The curve shows the characteristic plateaus indicative of hierarchical relaxation, as in the East model and the TPM.
Inset: average relaxation time as a function of inverse temperature, showing super-Arrhenius behaviour.}
\label{fig-spym-aging}
\end{figure}

This one-to-one correspondence between spin and defect configurations means that the thermodynamics of the SPyM is that of a free binary gas of plaquettes.  Furthermore, the relaxational dynamics is similar to that of a (three-dimensional) East model.  Figure~\ref{fig-spym-aging} shows the decay of the energy at low temperatures starting from a $T=\infty$ configuration.  We see the characteristic hierarchical decay of both the East model and the TPM:
the energy decays in steps with characteristic time scales $\tau_n = e^{n\beta J}$ with $n=0,1,2,\dots$.  These steps become apparent when plotting
as a function of the rescaled time variable $(T/J) \log t$~\cite{Sollich1999}.
The inset to Fig.~\ref{fig-spym-aging} shows that the equilibrium relaxation time of the SPyM is super-Arrhenius, as in the East model and the TPM.

\subsection{Phase transition in (annealed) coupled replicas}

The SPyM possesses the exact duality described in Sect.\ II.C. In particular, the properties of the SPyM in a field were studied in Ref.\ \cite{Heringa1989} (``model 1'' of that paper), where it was found that on the self-dual line there is a first-order transition between phases of small and large magnetisation terminating at a critical point in the $3D$ Ising class.  From those results we can directly infer the phase diagram of the two coupled SPyMs via the mapping of Sect.\ II.C.  The result is shown in Fig.~\ref{fig-eps-phase}.  This phase diagram is similar to that of the TPM, except that the range of phase coexistence is larger and the critical point occurs at higher temperature. 

\subsection{Evidence for a dynamical (space-time) phase transition}
\label{sec:spym-e-ens}

\begin{figure}
\includegraphics[width=\columnwidth]{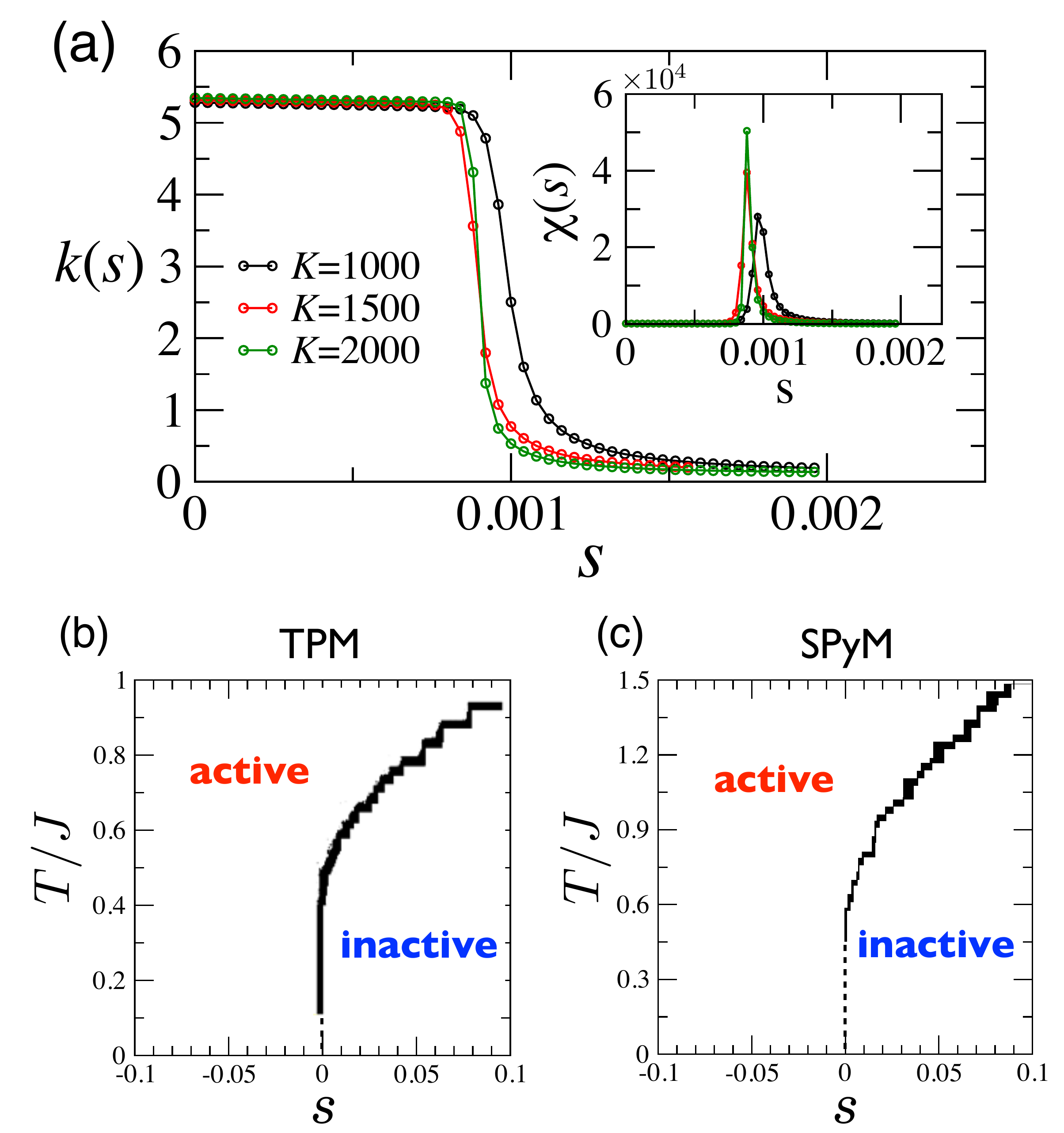}
\caption{(Color online)
(a) Activity as a function of $s$ in the SPyM for system size $L=4$, various trajectory lengths, and $T/J=0.65$.  The inset shows the corresponding susceptibility.  (b-c) $s$-ensemble phase diagrams for the TPM and SPyM.  The full curves are an estimate of the transition point from the simulations.  The dashed lines are extrapolations in the low temperature regime inaccessible to numerics.}
\label{fig-spym-s-ens}
\end{figure}

{As well as the phase transition for coupled replicas in the SPyM, we also present evidence for a space-time phase transition,
similar to that shown for the TPM in Fig.~\ref{fig-tpm-s-ens}.  The results for the SPyM are shown in Fig.~\ref{fig-spym-s-ens}, for temperature $T/J=0.65$ and linear
system size $L=4$.  There is good evidence for a sharp transition at $s=s^*>0$, as found in the TPM.}
%For the smaller system, the crossover in the activity is rather weak, but the data at $L=8$ show good evidence
%for a sharp transition at positive $s$, as found in the TPM.

In Fig.~\ref{fig-spym-s-ens}(b,c), we show how the crossover in activity varies with the temperature $T/J$, for both the TPM and the SPyM.
Simple estimates~\cite{Jack2011,Jack2013} indicate that if the inactive state is metastable and relaxes to equilibrium via some kind of nucleation process with
rate $\gamma_{\rm nuc}$ per unit volume, then $s^* \approx \gamma_{\rm nuc}/\delta k$, 
where $\delta k$ is the activity difference (per unit space-time) between the active and inactive states~\cite{Jack2011}.  We attribute the existence
of the transition in this model to a stable inactive state with almost no defects.  We expect that the rate for relaxation back
to equilibrium is a strongly decreasing function of temperature, which is consistent with the increasing $s^*$ as $T/J$ increases.
(The activity difference $\Delta k$ between active and inactive states increases with $T/J$ but this dependence is much weaker than that
of the relaxation rate.)

\section{Discussion}
\label{sec:discuss}

%
%(1) Both transitions are driven by transitions to metastable states with almost no defects.  It seems reasonable 
%that the existence of stable inactive states should be characteristic features of glassy systems, and it seems that both
%(annealed) coupled replicas and the $s$-ens are good ways to probe these states.  Ref Monasson-95 which says that
%for high overlap in annealed replicas, we samples states at an effective temperature $T/2$.
%
%(2) More precise sonnection between a stack of coupled replicas and the $s$-ensemble.  In one case think of a transfer matrix
%along the stack, in the other case think of the biased master operator (generator).  Both these ensembles have the same qualitative
%features and the same symmetries, reasonable to suppose they support the same kinds of phase transition.  For just two coupled replicas, the
%transition is different in detail but maybe similar physics?  
%
%(3) They key physical question wrt whether coupled replicas are relevant for dynamics
%seems to be: is the overlap a ``good order param'' for relaxation ??  (For East etc,
%transitiions between pairs configs with high overlap may still require convoluted dynamical pathways.  This should not be the case for plaq
%models etc)
%
%(4) To the extent that Franz-Parisi mean-field theory is relevant for plaq models (although there is no $T_K$),  does this have implications 
%for faciltiation theory vs Adam-Gibbs etc?

\subsection{Connection of phase transitions to long-lived metastable states}

{
The phase behaviour shown in Figs.~\ref{fig-eps-phase} and~\ref{fig-spym-s-ens} reveals striking similarities between thermodynamic transitions (for coupled
replicas) and dynamic transitions (based on dynamical activity).  We now argue that these transitions are connected to the existence
of long-lived metastable states, which are intrinsically linked to the glassy behaviour in these systems.  
%Our argument is based on the simple
%idea that if a state is metastable then small fluctuations within the state do not spread, since their spreading would lead to relaxation back to equilibrium.

%The lines $\eps=0$ and $s=0$ in Figs.~\ref{fig-eps-phase} and~\ref{fig-spym-s-ens} correspond to equilibrium
%behavior.  In equilibrium, at low temperature, the defect density is $N_d/N \approx e^{-\beta J}$. 
%In contrast, both the inactive state of 
%Fig.~\ref{fig-spym-s-ens} and the high-overlap state of Fig.~\ref{fig-eps-phase} have much lower defect concentrations: for large $s$, the system minimises its propensity for dynamical activity by removing defects, so that 
%$N_d/N\approx 0$ in this case; while for large coupling $\eps$ it is easy to show that $N_d/N \approx e^{-2\beta J}$
%since the system has an effective temperature $1/(2\beta)$~\cite{Monasson1995}.  

Consider first transitions for annealed replicas.  If we work at the phase coexistence point, but within the high-overlap phase, the system occupies low-energy
 states. For $\eps=0$, these states would be metastable.
This metastablility means that when localised low-overlap regions are generated by thermal fluctuations, these regions tend to shrink,
just as small fluctuations tend to shrink within classical nucleation theory.   Escape from the metastable state requires a collective process that
operates on some finite length scale.  As $\eps$ is increased from zero, this length scale increases, as does the associated free energy barrier: both diverge
at the coexistence point where the high-overlap phase becomes stable.

The situation for dynamical phase transitions is similar, except that one should think of trajectories of the system as $(d+1)$-dimensional objects that exist in space-time.  If one works at the dynamical phase coexistence point (some $s=s^*>0$) then inactive trajectories dominate the $s$-ensemble.  During these trajectories, the system remains localised in low-energy metastable states, with small thermal fluctuations of the activity, associated with space-time ``bubbles''~\cite{Chandler2010}.  If the trajectory length $\tau$ is less than the time required for escape from the metastable state, similar inactive trajectories can be generated with unbiased ($s=0$)
dynamics, 
by taking initial conditions from low-energy metastable states.  Transformation of such a trajectory into a \emph{typical} equilibrium trajectory involves the introduction of an active space-time bubble, which subsequently grows to macroscopic size.  The connection with metastability arises because if one introduces a small active bubble within an inactive trajectory, one expects to incur a cost in probability  (if this were not the case, the state would not be metastable since it would readily relax back to equilibrium).  As in the case of overlap fluctuations, the critical bubble size and the probability barrier increase as $s$ is increased from zero,
diverging at the coexistence point.

We argue that this analogy between phase coexistence phenomena induced by $s$- and $\eps$-fields provides a qualitative explanation
of the similarity between Figs.~\ref{fig-eps-phase} and~\ref{fig-spym-s-ens},
 in that both are linked to the existence of metastable states
that can be observed in unbiased ($s=0=\eps$) systems. 
The relevant metastable states have low energy: both the inactive state of 
Fig.~\ref{fig-spym-s-ens} and the high-overlap state of Fig.~\ref{fig-eps-phase} have much lower defect concentrations than the equilbrium
average value $c\approx{\rm e}^{-\beta J}$. For large $s$, the system minimises its propensity for dynamical activity by removing defects, so that 
$N_d/N\approx 0$; for large coupling $\eps$ it is easy to show that $N_d/N \approx e^{-2\beta J}$
since the system has an effective temperature $1/(2\beta)$~\cite{Monasson1995}. 
%
% The surface tensions that we consider are costs for interfaces between different states, with different energy densities, etc.   

 Evidence for phase coexistence induced by $s$ and $\eps$-fields have both been presented in atomistic models~\cite{Hedges2009,Berthier2013}.  
By contrast, in kinetically constrained models, dynamical phase coexistence can be induced by the $s$-field~\cite{Garrahan2007}
but there is no such transition as a function of $\eps$.  Metastable states do exist in these systems~\cite{Jack2013}, so phase coexistence at positive $s$ may
be expected, as argued above.  However, this metastability does not lead to a phase coexistence for any positive $\eps$ -- the reason is that the ``dynamical'' metastability that occurs in KCMs appears because thermal noise only generates a certain subset of the overlap fluctuations that are possible within the metastable state.  If all fluctuations in the overlap are possible, one finds that localised low-overlap regions grow easily in these systems. This means that there is no barrier between active and inactive states in the coupled-replica construction.  It is only within the dynamical $s$-ensemble that the metastability becomes apparent, due the restriction on the kinds of fluctuation that are generated by the thermal noise.

Finally, it is important to note that the inactive and high-overlap states in Figs.~\ref{fig-eps-phase} and~\ref{fig-spym-s-ens} are structurally distinct from their equilbrium
counterparts. 
% The
%ensembles that we consider are defined such that the system can sample metastable states that are not at all typical of equilibrium: we expect to find
%a transition whenever there exists a state whose surface tension with the equilibrium state is finite.  
This is quite different from ensembles with a 
quenched coupling between replicas~\cite{Berthier2015}, where the structures of high- and low-overlap states should be statistically (almost) indistinguishable.
}

%We also comment here on kinetically constrained models, in which overlap fluctuations are trivial, and there is no `static' surface tension.
%However, metastable states (and associated space-time surface tensions) 
%do exist in these models: this is possible because time-evolution between two states with high overlap 
%might require a very complicated (slow) dynamical process, due to the kinetic constraint.  In this case, a metastable state need not be stable against
%small perturbations in the overlap, because these perturbations are not generated by natural (finite-temperature) dynamical fluctuations.  Thus, for kinetically constrained models, the dynamical
%ensemble (\ref{s-ens}) includes active/inactive phase transitions, but the coupled replica systems do not.

\begin{figure}[t!]
\includegraphics[width=0.7\columnwidth]{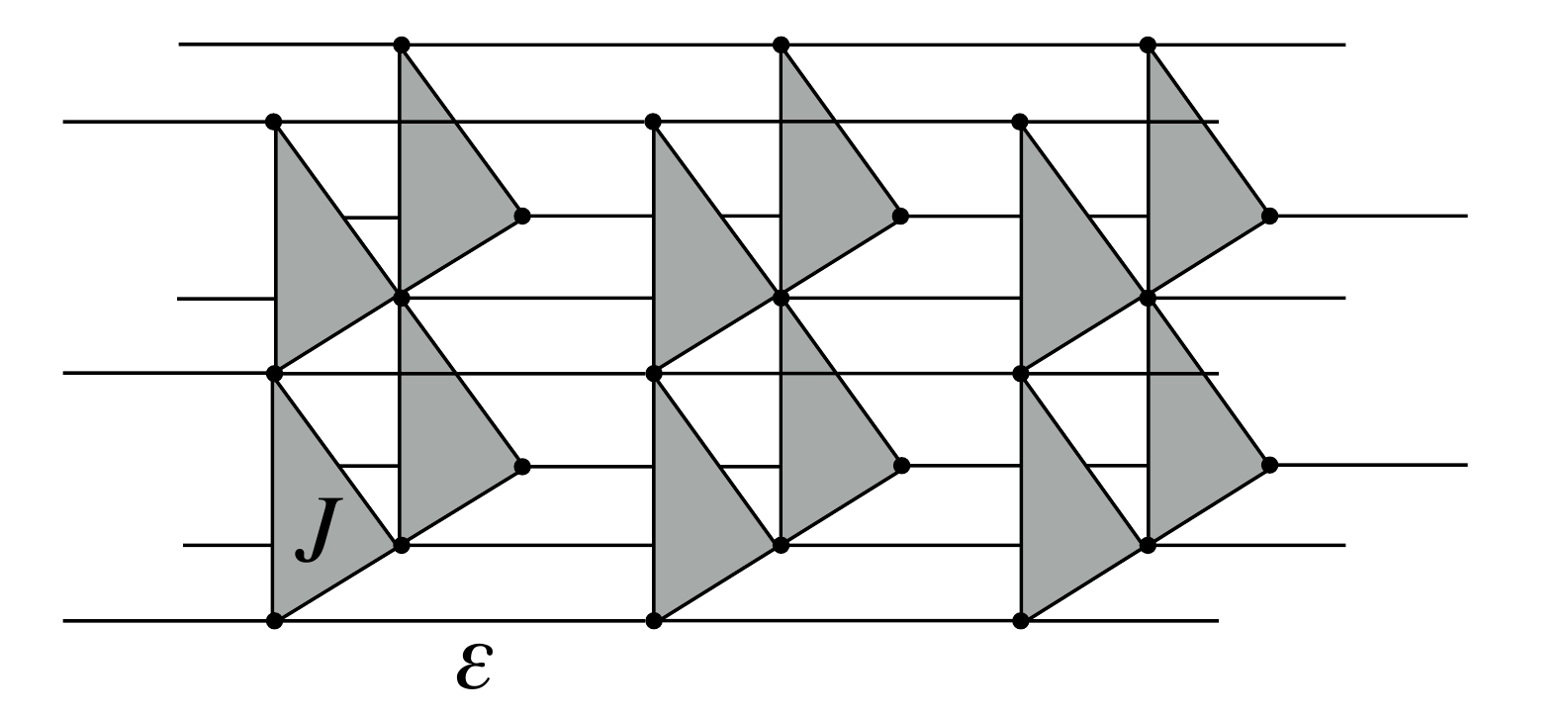}
\caption{Three dimensional stack of coupled two-dimensional TPMs. }
\label{fig-many-replicas}
\end{figure}

\subsection{Connection between multiple coupled replicas and biased-activity ensembles}

The connection between Figs.~\ref{fig-eps-phase} and~\ref{fig-spym-s-ens}
can be further motivated through a generalisation of the coupled two-replica system discussed above.  Given a plaquette model in dimension $d$, consider the associated $d+1$ system composed of many replicas of the $d$-dimensional system arranged parallel to each other along the extra dimension, see Fig.\ \ref{fig-many-replicas}.  Such system of $n$ coupled replicas has an energy
\begin{eqnarray}
E_{n}(\sigma^{1},\sigma^{2},\ldots,) &\equiv& E_J(\sigma^{1}) + E_J(\sigma^{2}) + \cdots
\nonumber \\
&& - \varepsilon \sum_{i} \left( \sigma_{i}^{1} \sigma_{i}^{2} + 
\sigma_{i}^{2} \sigma_{i}^{3} + \cdots \right)
.
\label{En}
\end{eqnarray}
Using the methods of Refs.\ \cite{Garrahan2014,Sasa2010,Deng2010} it is easy to prove that the partition sum of the $(d+1)$-dimensional problem also has an exact duality:
\begin{equation}
Z_n(J,\varepsilon) = (\sinh\beta J \sinh \beta \varepsilon)^{N n} Z_n(J^*,\varepsilon^*) ,
\label{Zn}
\end{equation}
where we have assumed periodic boundary conditions in the transverse direction, and $(J,\varepsilon)$ and $(J^*,\varepsilon^*)$ are related again by \eqref{Z2d2}. Similar results were found in \cite{Xu2004,*Xu2005} for other classes of plaquette models. Given this duality we expect the phenomenology of this many-replica system to be similar to that of two replicas, except that any phase transitions should be in the $(d+1)$-dimensional Ising universality class (assuming that both longitudinal and transverse dimensions are taken to infinity in the thermodynamic limit).

The partition sum $Z_{n}$ has a natural transfer matrix representation in the transverse direction, $Z_{n} = {\rm Tr}(\mathbb T^{n})$, with
\begin{eqnarray}
\mathbb T &=& \cosh^{N}(\beta J/2) e^{N \beta \epsilon} \bigotimes_{\mu} \left[ 1 +  \tanh(\beta J/2) 
\sigma_{i_{\mu}}^{z} \sigma_{j_{\mu}}^{z} \cdots \sigma_{k_{\mu}}^{z} \right]
\nonumber \\
&& 
\bigotimes_{i} \left( 1 +  e^{-2 \beta \epsilon} \sigma_{i}^{x} \right) ,
\end{eqnarray}
where $\sigma^{x,z}$ are Pauli matrices.  This in turn can be related to the generator of (imaginary time) quantum evolution in the usual manner \cite{Xu2004,*Xu2005} when the transverse coupling is large and the longitudinal one small [$e^{-2 \beta \varepsilon},\tanh(\beta J/2) \ll 1$], so that $Z_{n} \propto \exp \left( - t H \right)$, with 
\begin{equation}
H \equiv - h \sum_{i} \sigma_{i}^{x} - g \sum_{\mu}
\sigma_{i_{\mu}}^{z} \sigma_{j_{\mu}}^{z} \cdots \sigma_{k_{\mu}}^{z} ,
\label{HH}
\end{equation}
where
\begin{equation}
\delta t ~ h = e^{-2 \beta \varepsilon} , \;\;\;
\delta t ~ g = \beta J/2 , \;\;\;
t = n \delta t .
\end{equation}
The Hamiltonian (\ref{HH}) generates dynamics in the transverse direction.  While it is not derived from a stochastic operator it has the basic features of the generator \cite{Garrahan2007,Lecomte2007} for the dynamical ensemble defined in (\ref{s-ens}): an off-diagonal part (the $\sigma^{x}$ terms) that perform configuration changes and a diagonal part (the $\sigma^{z}$) plaquette terms associated to the escape rate.  The parameter $s$ in the $s$-ensemble operator controls the relative strength of the diagonal and off-diagonal terms \cite{Garrahan2007,Lecomte2007}, in analogy with the balance between $h$ and $g$ in \eqref{HH}.  Furthermore, the duality (\ref{Zn}) implies a duality $h \leftrightarrow g$ in (\ref{HH}), with the possibility of a dynamical transition at that self-dual point $g=h$.  This connection between a static transition in the $d+1$-dimensional problem \eqref{En} (itself closely connected to the static transition in the two-replica plaquette system) and a dynamical transition in the $d$-dimensional system \eqref{HH} provide another rationalisation of the similarities between Figs.~\ref{fig-eps-phase} and~\ref{fig-spym-s-ens}.

\subsection{Outlook}

Plaquette spin models have several features that make them attractive for studies of the glass transition.  As we have shown here, 
exact results can be derived, which guide numerical studies of phase behaviour and many-body correlations.  The models are also
computationally much less demanding than atomistic models of supercooled liquids, so that (for example) finite size scaling over a large range
of system sizes can be performed, to analyse phase transitions.  The equilibrium relaxation of the models follows a dynamical facilitation scenario, in which point defects
play a central role.  However, there are strong many-body correlations, and the statistics of overlap fluctuations are rich and complex, as 
anticipated in the theory of Franz and Parisi~\cite{Franz1997}.  
In this sense, the models provide a bridge between different theories.  Indeed, as argued in~\cite{Jack2005caging}, 
one might describe plaquette models by a modified form of RFOT, but with two important caveats: (i) the analogue of the Kauzmann transition occurs
at zero temperature in these models (ii) the interfacial cost associated with growing droplets of a new state within a typical equilibrium state scales
logarithmically in the droplet size (not as a power law as anticipated by RFOT).

Looking forward, we hope that further work on plaquette models (particularly in $d=3$) will show to what extent mean-field~\cite{Franz1997} and RFOT ideas can be
modified to apply in this setting.  We can imagine that the apparently different physical pictures envisaged by thermodynamic and dynamical theories of the glass transition~\cite{Biroli2013} might \emph{both} be applicable in these models.  In that case, it is not clear whether some new results would be required to discriminate between
the theories, or whether they might in fact offer complementary descriptions of the same phenomena.

\bigskip

\noindent
{\bf Acknowledgements.} We thank Nigel Wilding for helpful discussions and for providing the 
Ising model data for Fig.~\ref{fig-tpm-field}. This work was supported in part by EPSRC Grant No. EP/ I017828/1.  RLJ was supported by the EPRSC through grant EP/I003797/1.


\begin{thebibliography}{53}%
\makeatletter
\providecommand \@ifxundefined [1]{%
 \@ifx{#1\undefined}
}%
\providecommand \@ifnum [1]{%
 \ifnum #1\expandafter \@firstoftwo
 \else \expandafter \@secondoftwo
 \fi
}%
\providecommand \@ifx [1]{%
 \ifx #1\expandafter \@firstoftwo
 \else \expandafter \@secondoftwo
 \fi
}%
\providecommand \natexlab [1]{#1}%
\providecommand \enquote  [1]{``#1''}%
\providecommand \bibnamefont  [1]{#1}%
\providecommand \bibfnamefont [1]{#1}%
\providecommand \citenamefont [1]{#1}%
\providecommand \href@noop [0]{\@secondoftwo}%
\providecommand \href [0]{\begingroup \@sanitize@url \@href}%
\providecommand \@href[1]{\@@startlink{#1}\@@href}%
\providecommand \@@href[1]{\endgroup#1\@@endlink}%
\providecommand \@sanitize@url [0]{\catcode `\\12\catcode `\$12\catcode
  `\&12\catcode `\#12\catcode `\^12\catcode `\_12\catcode `\%12\relax}%
\providecommand \@@startlink[1]{}%
\providecommand \@@endlink[0]{}%
\providecommand \url  [0]{\begingroup\@sanitize@url \@url }%
\providecommand \@url [1]{\endgroup\@href {#1}{\urlprefix }}%
\providecommand \urlprefix  [0]{URL }%
\providecommand \Eprint [0]{\href }%
\providecommand \doibase [0]{http://dx.doi.org/}%
\providecommand \selectlanguage [0]{\@gobble}%
\providecommand \bibinfo  [0]{\@secondoftwo}%
\providecommand \bibfield  [0]{\@secondoftwo}%
\providecommand \translation [1]{[#1]}%
\providecommand \BibitemOpen [0]{}%
\providecommand \bibitemStop [0]{}%
\providecommand \bibitemNoStop [0]{.\EOS\space}%
\providecommand \EOS [0]{\spacefactor3000\relax}%
\providecommand \BibitemShut  [1]{\csname bibitem#1\endcsname}%
\let\auto@bib@innerbib\@empty
%</preamble>
\bibitem [{\citenamefont {Ediger}\ \emph {et~al.}(1996)\citenamefont {Ediger},
  \citenamefont {Angell},\ and\ \citenamefont {Nagel}}]{Ediger1996}%
  \BibitemOpen
  \bibfield  {author} {\bibinfo {author} {\bibfnamefont {M.~D.}\ \bibnamefont
  {Ediger}}, \bibinfo {author} {\bibfnamefont {C.~A.}\ \bibnamefont {Angell}},
  \ and\ \bibinfo {author} {\bibfnamefont {S.~R.}\ \bibnamefont {Nagel}},\
  }\href@noop {} {\bibfield  {journal} {\bibinfo  {journal} {J. Phys. Chem.}\
  }\textbf {\bibinfo {volume} {100}},\ \bibinfo {pages} {13200} (\bibinfo
  {year} {1996})}\BibitemShut {NoStop}%
\bibitem [{\citenamefont {Cavagna}(2009)}]{Cavagna2009}%
  \BibitemOpen
  \bibfield  {author} {\bibinfo {author} {\bibfnamefont {A.}~\bibnamefont
  {Cavagna}},\ }\href@noop {} {\bibfield  {journal} {\bibinfo  {journal} {Phys.
  Rep.}\ }\textbf {\bibinfo {volume} {476}},\ \bibinfo {pages} {51} (\bibinfo
  {year} {2009})}\BibitemShut {NoStop}%
\bibitem [{\citenamefont {Berthier}\ and\ \citenamefont
  {Biroli}(2011)}]{Berthier2011}%
  \BibitemOpen
  \bibfield  {author} {\bibinfo {author} {\bibfnamefont {L.}~\bibnamefont
  {Berthier}}\ and\ \bibinfo {author} {\bibfnamefont {G.}~\bibnamefont
  {Biroli}},\ }\href {\doibase 10.1103/RevModPhys.83.587} {\bibfield  {journal}
  {\bibinfo  {journal} {Rev. Mod. Phys.}\ }\textbf {\bibinfo {volume} {83}},\
  \bibinfo {pages} {587} (\bibinfo {year} {2011})}\BibitemShut {NoStop}%
\bibitem [{\citenamefont {Biroli}\ and\ \citenamefont
  {Garrahan}(2013)}]{Biroli2013}%
  \BibitemOpen
  \bibfield  {author} {\bibinfo {author} {\bibfnamefont {G.}~\bibnamefont
  {Biroli}}\ and\ \bibinfo {author} {\bibfnamefont {J.~P.}\ \bibnamefont
  {Garrahan}},\ }\href {\doibase 10.1063/1.4795539} {\bibfield  {journal}
  {\bibinfo  {journal} {J. Chem. Phys.}\ }\textbf {\bibinfo {volume} {138}},\
  \bibinfo {eid} {12A301} (\bibinfo {year} {2013})}\BibitemShut {NoStop}%
\bibitem [{\citenamefont {Lubchenko}\ and\ \citenamefont
  {Wolynes}(2007)}]{Lubchenko2007}%
  \BibitemOpen
  \bibfield  {author} {\bibinfo {author} {\bibfnamefont {V.}~\bibnamefont
  {Lubchenko}}\ and\ \bibinfo {author} {\bibfnamefont {P.~G.}\ \bibnamefont
  {Wolynes}},\ }\href@noop {} {\bibfield  {journal} {\bibinfo  {journal} {Annu.
  Rev. Phys. Chem.}\ }\textbf {\bibinfo {volume} {58}},\ \bibinfo {pages} {235}
  (\bibinfo {year} {2007})}\BibitemShut {NoStop}%
\bibitem [{\citenamefont {Franz}\ and\ \citenamefont
  {Parisi}(1997)}]{Franz1997}%
  \BibitemOpen
  \bibfield  {author} {\bibinfo {author} {\bibfnamefont {S.}~\bibnamefont
  {Franz}}\ and\ \bibinfo {author} {\bibfnamefont {G.}~\bibnamefont {Parisi}},\
  }\href {\doibase 10.1103/PhysRevLett.79.2486} {\bibfield  {journal} {\bibinfo
   {journal} {Phys. Rev. Lett.}\ }\textbf {\bibinfo {volume} {79}},\ \bibinfo
  {pages} {2486} (\bibinfo {year} {1997})}\BibitemShut {NoStop}%
\bibitem [{\citenamefont {Chandler}\ and\ \citenamefont
  {Garrahan}(2010)}]{Chandler2010}%
  \BibitemOpen
  \bibfield  {author} {\bibinfo {author} {\bibfnamefont {D.}~\bibnamefont
  {Chandler}}\ and\ \bibinfo {author} {\bibfnamefont {J.~P.}\ \bibnamefont
  {Garrahan}},\ }\href
  {http://dx.doi.org/10.1146/annurev.physchem.040808.090405} {\bibfield
  {journal} {\bibinfo  {journal} {Annu. Rev. Phys. Chem.}\ }\textbf {\bibinfo
  {volume} {61}},\ \bibinfo {pages} {191} (\bibinfo {year} {2010})}\BibitemShut
  {NoStop}%
\bibitem [{\citenamefont {Garrahan}\ \emph {et~al.}(2007)\citenamefont
  {Garrahan}, \citenamefont {Jack}, \citenamefont {Lecomte}, \citenamefont
  {Pitard}, \citenamefont {van Duijvendijk},\ and\ \citenamefont {van
  Wijland}}]{Garrahan2007}%
  \BibitemOpen
  \bibfield  {author} {\bibinfo {author} {\bibfnamefont {J.~P.}\ \bibnamefont
  {Garrahan}}, \bibinfo {author} {\bibfnamefont {R.~L.}\ \bibnamefont {Jack}},
  \bibinfo {author} {\bibfnamefont {V.}~\bibnamefont {Lecomte}}, \bibinfo
  {author} {\bibfnamefont {E.}~\bibnamefont {Pitard}}, \bibinfo {author}
  {\bibfnamefont {K.}~\bibnamefont {van Duijvendijk}}, \ and\ \bibinfo {author}
  {\bibfnamefont {F.}~\bibnamefont {van Wijland}},\ }\href {\doibase
  10.1103/PhysRevLett.98.195702} {\bibfield  {journal} {\bibinfo  {journal}
  {Phys. Rev. Lett.}\ }\textbf {\bibinfo {volume} {98}},\ \bibinfo {pages}
  {195702} (\bibinfo {year} {2007})}\BibitemShut {NoStop}%
\bibitem [{\citenamefont {Hedges}\ \emph {et~al.}(2009)\citenamefont {Hedges},
  \citenamefont {Jack}, \citenamefont {Garrahan},\ and\ \citenamefont
  {Chandler}}]{Hedges2009}%
  \BibitemOpen
  \bibfield  {author} {\bibinfo {author} {\bibfnamefont {L.~O.}\ \bibnamefont
  {Hedges}}, \bibinfo {author} {\bibfnamefont {R.~L.}\ \bibnamefont {Jack}},
  \bibinfo {author} {\bibfnamefont {J.~P.}\ \bibnamefont {Garrahan}}, \ and\
  \bibinfo {author} {\bibfnamefont {D.}~\bibnamefont {Chandler}},\ }\href
  {\doibase 10.1126/science.1166665} {\bibfield  {journal} {\bibinfo  {journal}
  {Science}\ }\textbf {\bibinfo {volume} {323}},\ \bibinfo {pages} {1309}
  (\bibinfo {year} {2009})}\BibitemShut {NoStop}%
\bibitem [{\citenamefont {Garrahan}(2002)}]{Garrahan2002}%
  \BibitemOpen
  \bibfield  {author} {\bibinfo {author} {\bibfnamefont {J.~P.}\ \bibnamefont
  {Garrahan}},\ }\href {\doibase PII S0953-8984(02)31606-0} {\bibfield
  {journal} {\bibinfo  {journal} {J. Phys. Condens. Matter}\ }\textbf {\bibinfo
  {volume} {14}},\ \bibinfo {pages} {1571} (\bibinfo {year}
  {2002})}\BibitemShut {NoStop}%
\bibitem [{\citenamefont {J\"{a}ckle}\ and\ \citenamefont
  {Eisinger}(1991)}]{Jackle1991}%
  \BibitemOpen
  \bibfield  {author} {\bibinfo {author} {\bibfnamefont {J.}~\bibnamefont
  {J\"{a}ckle}}\ and\ \bibinfo {author} {\bibfnamefont {S.}~\bibnamefont
  {Eisinger}},\ }\href {\doibase 10.1007/BF01453764} {\bibfield  {journal}
  {\bibinfo  {journal} {Z. Phys. B}\ }\textbf {\bibinfo {volume} {84}},\
  \bibinfo {pages} {115} (\bibinfo {year} {1991})}\BibitemShut {NoStop}%
\bibitem [{\citenamefont {Ritort}\ and\ \citenamefont
  {Sollich}(2003)}]{Ritort2003}%
  \BibitemOpen
  \bibfield  {author} {\bibinfo {author} {\bibfnamefont {F.}~\bibnamefont
  {Ritort}}\ and\ \bibinfo {author} {\bibfnamefont {P.}~\bibnamefont
  {Sollich}},\ }\href@noop {} {\bibfield  {journal} {\bibinfo  {journal} {Adv.
  Phys.}\ }\textbf {\bibinfo {volume} {52}},\ \bibinfo {pages} {219} (\bibinfo
  {year} {2003})}\BibitemShut {NoStop}%
\bibitem [{\citenamefont {Garrahan}\ and\ \citenamefont
  {Chandler}(2003)}]{Garrahan2003}%
  \BibitemOpen
  \bibfield  {author} {\bibinfo {author} {\bibfnamefont {J.}~\bibnamefont
  {Garrahan}}\ and\ \bibinfo {author} {\bibfnamefont {D.}~\bibnamefont
  {Chandler}},\ }\href {\doibase DOI 10.1073/pnas.1233719100} {\bibfield
  {journal} {\bibinfo  {journal} {Proc. Natl. Acad. Sci. USA}\ }\textbf
  {\bibinfo {volume} {100}},\ \bibinfo {pages} {9710} (\bibinfo {year}
  {2003})}\BibitemShut {NoStop}%
\bibitem [{\citenamefont {Newman}\ and\ \citenamefont
  {Moore}(1999)}]{Newman1999}%
  \BibitemOpen
  \bibfield  {author} {\bibinfo {author} {\bibfnamefont {M.}~\bibnamefont
  {Newman}}\ and\ \bibinfo {author} {\bibfnamefont {C.}~\bibnamefont {Moore}},\
  }\href@noop {} {\bibfield  {journal} {\bibinfo  {journal} {Phys. Rev. E}\
  }\textbf {\bibinfo {volume} {60}},\ \bibinfo {pages} {5068} (\bibinfo {year}
  {1999})}\BibitemShut {NoStop}%
\bibitem [{\citenamefont {Garrahan}\ and\ \citenamefont
  {Newman}(2000)}]{Garrahan2000}%
  \BibitemOpen
  \bibfield  {author} {\bibinfo {author} {\bibfnamefont {J.~P.}\ \bibnamefont
  {Garrahan}}\ and\ \bibinfo {author} {\bibfnamefont {M.}~\bibnamefont
  {Newman}},\ }\href@noop {} {\bibfield  {journal} {\bibinfo  {journal} {Phys.
  Rev. E}\ }\textbf {\bibinfo {volume} {62}},\ \bibinfo {pages} {7670}
  (\bibinfo {year} {2000})}\BibitemShut {NoStop}%
\bibitem [{\citenamefont {Heringa}\ \emph {et~al.}(1989)\citenamefont
  {Heringa}, \citenamefont {Bl\"ote},\ and\ \citenamefont
  {Hoogland}}]{Heringa1989}%
  \BibitemOpen
  \bibfield  {author} {\bibinfo {author} {\bibfnamefont {J.~R.}\ \bibnamefont
  {Heringa}}, \bibinfo {author} {\bibfnamefont {H.~W.~J.}\ \bibnamefont
  {Bl\"ote}}, \ and\ \bibinfo {author} {\bibfnamefont {A.}~\bibnamefont
  {Hoogland}},\ }\href {\doibase 10.1103/PhysRevLett.63.1546} {\bibfield
  {journal} {\bibinfo  {journal} {Phys. Rev. Lett.}\ }\textbf {\bibinfo
  {volume} {63}},\ \bibinfo {pages} {1546} (\bibinfo {year}
  {1989})}\BibitemShut {NoStop}%
\bibitem [{\citenamefont {Sasa}(2010)}]{Sasa2010}%
  \BibitemOpen
  \bibfield  {author} {\bibinfo {author} {\bibfnamefont {S.}~\bibnamefont
  {Sasa}},\ }\href@noop {} {\bibfield  {journal} {\bibinfo  {journal} {J. Phys.
  A}\ }\textbf {\bibinfo {volume} {43}},\ \bibinfo {pages} {465002} (\bibinfo
  {year} {2010})}\BibitemShut {NoStop}%
\bibitem [{\citenamefont {Garrahan}(2014)}]{Garrahan2014}%
  \BibitemOpen
  \bibfield  {author} {\bibinfo {author} {\bibfnamefont {J.~P.}\ \bibnamefont
  {Garrahan}},\ }\href {\doibase 10.1103/PhysRevE.89.030301} {\bibfield
  {journal} {\bibinfo  {journal} {Phys. Rev. E}\ }\textbf {\bibinfo {volume}
  {89}},\ \bibinfo {pages} {030301} (\bibinfo {year} {2014})}\BibitemShut
  {NoStop}%
\bibitem [{\citenamefont {Elmatad}\ \emph {et~al.}(2010)\citenamefont
  {Elmatad}, \citenamefont {Jack}, \citenamefont {Chandler},\ and\
  \citenamefont {Garrahan}}]{Elmatad2010}%
  \BibitemOpen
  \bibfield  {author} {\bibinfo {author} {\bibfnamefont {Y.~S.}\ \bibnamefont
  {Elmatad}}, \bibinfo {author} {\bibfnamefont {R.~L.}\ \bibnamefont {Jack}},
  \bibinfo {author} {\bibfnamefont {D.}~\bibnamefont {Chandler}}, \ and\
  \bibinfo {author} {\bibfnamefont {J.~P.}\ \bibnamefont {Garrahan}},\
  }\href@noop {} {\bibfield  {journal} {\bibinfo  {journal} {Proc. Natl. Acad.
  Sci. USA}\ }\textbf {\bibinfo {volume} {107}},\ \bibinfo {pages} {12793}
  (\bibinfo {year} {2010})}\BibitemShut {NoStop}%
\bibitem [{\citenamefont {Jack}\ and\ \citenamefont
  {Garrahan}(2005)}]{Jack2005caging}%
  \BibitemOpen
  \bibfield  {author} {\bibinfo {author} {\bibfnamefont {R.~L.}\ \bibnamefont
  {Jack}}\ and\ \bibinfo {author} {\bibfnamefont {J.~P.}\ \bibnamefont
  {Garrahan}},\ }\href@noop {} {\bibfield  {journal} {\bibinfo  {journal} {J.
  Chem. Phys.}\ }\textbf {\bibinfo {volume} {123}},\ \bibinfo {pages} {164508}
  (\bibinfo {year} {2005})}\BibitemShut {NoStop}%
\bibitem [{\citenamefont {Johnston}(2012)}]{Johnston2012}%
  \BibitemOpen
  \bibfield  {author} {\bibinfo {author} {\bibfnamefont {D.}~\bibnamefont
  {Johnston}},\ }\href@noop {} {\bibfield  {journal} {\bibinfo  {journal} {J.
  Phys. A}\ }\textbf {\bibinfo {volume} {45}},\ \bibinfo {pages} {405001}
  (\bibinfo {year} {2012})}\BibitemShut {NoStop}%
\bibitem [{\citenamefont {Foini}\ \emph {et~al.}(2012)\citenamefont {Foini},
  \citenamefont {Krzakala},\ and\ \citenamefont {Zamponi}}]{Foini2012}%
  \BibitemOpen
  \bibfield  {author} {\bibinfo {author} {\bibfnamefont {L.}~\bibnamefont
  {Foini}}, \bibinfo {author} {\bibfnamefont {F.}~\bibnamefont {Krzakala}}, \
  and\ \bibinfo {author} {\bibfnamefont {F.}~\bibnamefont {Zamponi}},\
  }\href@noop {} {\bibfield  {journal} {\bibinfo  {journal} {J. Stat. Mech.}\
  ,\ \bibinfo {pages} {P06013}} (\bibinfo {year} {2012})}\BibitemShut {NoStop}%
\bibitem [{\citenamefont {Franz}\ and\ \citenamefont {Parisi}()}]{Franz2013}%
  \BibitemOpen
  \bibfield  {author} {\bibinfo {author} {\bibfnamefont {S.}~\bibnamefont
  {Franz}}\ and\ \bibinfo {author} {\bibfnamefont {G.}~\bibnamefont {Parisi}},\
  }\href@noop {} {\bibinfo  {journal} {arXiv:1307.4955}\ }\BibitemShut
  {NoStop}%
\bibitem [{\citenamefont {Berthier}(2013)}]{Berthier2013}%
  \BibitemOpen
\bibfield  {journal} {  }\bibfield  {author} {\bibinfo {author} {\bibfnamefont
  {L.}~\bibnamefont {Berthier}},\ }\href {\doibase 10.1103/PhysRevE.88.022313}
  {\bibfield  {journal} {\bibinfo  {journal} {Phys. Rev. E}\ }\textbf {\bibinfo
  {volume} {88}},\ \bibinfo {pages} {022313} (\bibinfo {year}
  {2013})}\BibitemShut {NoStop}%
\bibitem [{\citenamefont {Parisi}\ and\ \citenamefont {Seoane}()}]{Parisi2013}%
  \BibitemOpen
  \bibfield  {author} {\bibinfo {author} {\bibfnamefont {G.}~\bibnamefont
  {Parisi}}\ and\ \bibinfo {author} {\bibfnamefont {B.}~\bibnamefont
  {Seoane}},\ }\href@noop {} {\bibinfo  {journal} {arXiv:1311.1465}\
  }\BibitemShut {NoStop}%
\bibitem [{\citenamefont {Nicolaides}\ and\ \citenamefont
  {Bruce}(1988)}]{Nico1988}%
  \BibitemOpen
\bibfield  {journal} {  }\bibfield  {author} {\bibinfo {author} {\bibfnamefont
  {D.}~\bibnamefont {Nicolaides}}\ and\ \bibinfo {author} {\bibfnamefont
  {A.~D.}\ \bibnamefont {Bruce}},\ }\href@noop {} {\bibfield  {journal}
  {\bibinfo  {journal} {J. Phys. A}\ }\textbf {\bibinfo {volume} {21}},\
  \bibinfo {pages} {233} (\bibinfo {year} {1988})}\BibitemShut {NoStop}%
\bibitem [{\citenamefont {Wilding}\ and\ \citenamefont
  {Mueller}(1995)}]{Wilding1995}%
  \BibitemOpen
  \bibfield  {author} {\bibinfo {author} {\bibfnamefont {N.~B.}\ \bibnamefont
  {Wilding}}\ and\ \bibinfo {author} {\bibfnamefont {M.}~\bibnamefont
  {Mueller}},\ }\href@noop {} {\bibfield  {journal} {\bibinfo  {journal} {J.
  Chem. Phys.}\ }\textbf {\bibinfo {volume} {102}},\ \bibinfo {pages} {2562}
  (\bibinfo {year} {1995})}\BibitemShut {NoStop}%
\bibitem [{\citenamefont {Bortz}\ \emph {et~al.}(1975)\citenamefont {Bortz},
  \citenamefont {Kalos},\ and\ \citenamefont {Lebowitz}}]{Bortz1975}%
  \BibitemOpen
  \bibfield  {author} {\bibinfo {author} {\bibfnamefont {A.~B.}\ \bibnamefont
  {Bortz}}, \bibinfo {author} {\bibfnamefont {M.~H.}\ \bibnamefont {Kalos}}, \
  and\ \bibinfo {author} {\bibfnamefont {J.~L.}\ \bibnamefont {Lebowitz}},\
  }\href@noop {} {\bibfield  {journal} {\bibinfo  {journal} {J. Comp. Phys.}\
  }\textbf {\bibinfo {volume} {17}},\ \bibinfo {pages} {10} (\bibinfo {year}
  {1975})}\BibitemShut {NoStop}%
\bibitem [{\citenamefont {Newman}\ and\ \citenamefont
  {Barkema}(1999)}]{NewmanBook}%
  \BibitemOpen
  \bibfield  {author} {\bibinfo {author} {\bibfnamefont {M.~E.~J.}\
  \bibnamefont {Newman}}\ and\ \bibinfo {author} {\bibfnamefont {G.~T.}\
  \bibnamefont {Barkema}},\ }\href@noop {} {\emph {\bibinfo {title} {Monte
  Carlo Methods in Statistical Physics}}}\ (\bibinfo  {publisher} {OUP,
  Oxford},\ \bibinfo {year} {1999})\BibitemShut {NoStop}%
\bibitem [{\citenamefont {Bruce}\ and\ \citenamefont
  {Wilding}(2003)}]{bruce2003}%
  \BibitemOpen
  \bibfield  {author} {\bibinfo {author} {\bibfnamefont {A.~D.}\ \bibnamefont
  {Bruce}}\ and\ \bibinfo {author} {\bibfnamefont {N.~B.}\ \bibnamefont
  {Wilding}},\ }\href@noop {} {\bibfield  {journal} {\bibinfo  {journal} {Adv.
  Chem. Phys.}\ }\textbf {\bibinfo {volume} {127}},\ \bibinfo {pages} {1}
  (\bibinfo {year} {2003})}\BibitemShut {NoStop}%
\bibitem [{\citenamefont {Fredrickson}\ and\ \citenamefont
  {Andersen}(1984)}]{Fredrickson1984}%
  \BibitemOpen
  \bibfield  {author} {\bibinfo {author} {\bibfnamefont {G.~H.}\ \bibnamefont
  {Fredrickson}}\ and\ \bibinfo {author} {\bibfnamefont {H.~C.}\ \bibnamefont
  {Andersen}},\ }\href@noop {} {\bibfield  {journal} {\bibinfo  {journal}
  {Phys. Rev. Lett.}\ }\textbf {\bibinfo {volume} {53}},\ \bibinfo {pages}
  {1244} (\bibinfo {year} {1984})}\BibitemShut {NoStop}%
\bibitem [{\citenamefont {Lecomte}\ \emph {et~al.}(2007)\citenamefont
  {Lecomte}, \citenamefont {Appert-Rolland},\ and\ \citenamefont {van
  Wijland}}]{Lecomte2007}%
  \BibitemOpen
  \bibfield  {author} {\bibinfo {author} {\bibfnamefont {V.}~\bibnamefont
  {Lecomte}}, \bibinfo {author} {\bibfnamefont {C.}~\bibnamefont
  {Appert-Rolland}}, \ and\ \bibinfo {author} {\bibfnamefont {F.}~\bibnamefont
  {van Wijland}},\ }\href {\doibase 10.1007/s10955-006-9254-0} {\bibfield
  {journal} {\bibinfo  {journal} {J. Stat. Phys.}\ }\textbf {\bibinfo {volume}
  {127}},\ \bibinfo {pages} {51} (\bibinfo {year} {2007})}\BibitemShut
  {NoStop}%
\bibitem [{\citenamefont {Baiesi}\ \emph {et~al.}(2009)\citenamefont {Baiesi},
  \citenamefont {Maes},\ and\ \citenamefont {Wynants}}]{Baiesi2009}%
  \BibitemOpen
  \bibfield  {author} {\bibinfo {author} {\bibfnamefont {M.}~\bibnamefont
  {Baiesi}}, \bibinfo {author} {\bibfnamefont {C.}~\bibnamefont {Maes}}, \ and\
  \bibinfo {author} {\bibfnamefont {B.}~\bibnamefont {Wynants}},\ }\href@noop
  {} {\bibfield  {journal} {\bibinfo  {journal} {Phys. Rev. Lett.}\ }\textbf
  {\bibinfo {volume} {103}},\ \bibinfo {pages} {010602} (\bibinfo {year}
  {2009})}\BibitemShut {NoStop}%
\bibitem [{\citenamefont {Touchette}(2009)}]{Touchette2009}%
  \BibitemOpen
  \bibfield  {author} {\bibinfo {author} {\bibfnamefont {H.}~\bibnamefont
  {Touchette}},\ }\href {\doibase 10.1016/j.physrep.2009.05.002} {\bibfield
  {journal} {\bibinfo  {journal} {Phys. Rep.}\ }\textbf {\bibinfo {volume}
  {478}},\ \bibinfo {pages} {1} (\bibinfo {year} {2009})}\BibitemShut {NoStop}%
\bibitem [{\citenamefont {Speck}\ \emph {et~al.}(2012)\citenamefont {Speck},
  \citenamefont {Malins},\ and\ \citenamefont {Royall}}]{Speck2012}%
  \BibitemOpen
  \bibfield  {author} {\bibinfo {author} {\bibfnamefont {T.}~\bibnamefont
  {Speck}}, \bibinfo {author} {\bibfnamefont {A.}~\bibnamefont {Malins}}, \
  and\ \bibinfo {author} {\bibfnamefont {C.~P.}\ \bibnamefont {Royall}},\
  }\href@noop {} {\bibfield  {journal} {\bibinfo  {journal} {Phys. Rev. Lett.}\
  }\textbf {\bibinfo {volume} {109}},\ \bibinfo {pages} {195703} (\bibinfo
  {year} {2012})}\BibitemShut {NoStop}%
\bibitem [{\citenamefont {Elmatad}\ \emph {et~al.}(2009)\citenamefont
  {Elmatad}, \citenamefont {Chandler},\ and\ \citenamefont
  {Garrahan}}]{Elmatad2009}%
  \BibitemOpen
  \bibfield  {author} {\bibinfo {author} {\bibfnamefont {Y.~S.}\ \bibnamefont
  {Elmatad}}, \bibinfo {author} {\bibfnamefont {D.}~\bibnamefont {Chandler}}, \
  and\ \bibinfo {author} {\bibfnamefont {J.~P.}\ \bibnamefont {Garrahan}},\
  }\href {http://pubs.acs.org/doi/abs/10.1021/jp810362g} {\bibfield  {journal}
  {\bibinfo  {journal} {J. Phys. Chem. B}\ }\textbf {\bibinfo {volume} {113}},\
  \bibinfo {pages} {5563} (\bibinfo {year} {2009})}\BibitemShut {NoStop}%
\bibitem [{\citenamefont {Bolhuis}\ \emph {et~al.}(2002)\citenamefont
  {Bolhuis}, \citenamefont {Chandler}, \citenamefont {Dellago},\ and\
  \citenamefont {Geissler}}]{Bolhuis2002}%
  \BibitemOpen
  \bibfield  {author} {\bibinfo {author} {\bibfnamefont {P.~G.}\ \bibnamefont
  {Bolhuis}}, \bibinfo {author} {\bibfnamefont {D.}~\bibnamefont {Chandler}},
  \bibinfo {author} {\bibfnamefont {C.}~\bibnamefont {Dellago}}, \ and\
  \bibinfo {author} {\bibfnamefont {P.~L.}\ \bibnamefont {Geissler}},\
  }\href@noop {} {\bibfield  {journal} {\bibinfo  {journal} {Annu. Rev. Phys.
  Chem.}\ }\textbf {\bibinfo {volume} {53}},\ \bibinfo {pages} {291} (\bibinfo
  {year} {2002})}\BibitemShut {NoStop}%
\bibitem [{\citenamefont {Chetrite}\ and\ \citenamefont
  {Touchette}(2013)}]{Chetrite2013}%
  \BibitemOpen
  \bibfield  {author} {\bibinfo {author} {\bibfnamefont {R.}~\bibnamefont
  {Chetrite}}\ and\ \bibinfo {author} {\bibfnamefont {H.}~\bibnamefont
  {Touchette}},\ }\href@noop {} {\bibfield  {journal} {\bibinfo  {journal}
  {Phys. Rev. Lett.}\ }\textbf {\bibinfo {volume} {111}},\ \bibinfo {pages}
  {120601} (\bibinfo {year} {2013})}\BibitemShut {NoStop}%
\bibitem [{\citenamefont {Budini}\ \emph {et~al.}(2014)\citenamefont {Budini},
  \citenamefont {Turner},\ and\ \citenamefont {Garrahan}}]{Budini2014}%
  \BibitemOpen
  \bibfield  {author} {\bibinfo {author} {\bibfnamefont {A.~A.}\ \bibnamefont
  {Budini}}, \bibinfo {author} {\bibfnamefont {R.~M.}\ \bibnamefont {Turner}},
  \ and\ \bibinfo {author} {\bibfnamefont {J.~P.}\ \bibnamefont {Garrahan}},\
  }\href {http://stacks.iop.org/1742-5468/2014/i=3/a=P03012} {\bibfield
  {journal} {\bibinfo  {journal} {J. Stat. Mech.}\ ,\ \bibinfo {pages}
  {P03012}} (\bibinfo {year} {2014})}\BibitemShut {NoStop}%
\bibitem [{Kiu()}]{Kiukas2015}%
  \BibitemOpen
  \href@noop {} {}\bibinfo {note} {J.~Kiukas, M.~Guta, I.~Lesanovsky and
  J.~P.~Garrahan, arXiv:1503.05716}\BibitemShut {NoStop}%
\bibitem [{\citenamefont {Turner}\ \emph {et~al.}(2014)\citenamefont {Turner},
  \citenamefont {Speck},\ and\ \citenamefont {Garrahan}}]{Turner2014}%
  \BibitemOpen
  \bibfield  {author} {\bibinfo {author} {\bibfnamefont {R.~M.}\ \bibnamefont
  {Turner}}, \bibinfo {author} {\bibfnamefont {T.}~\bibnamefont {Speck}}, \
  and\ \bibinfo {author} {\bibfnamefont {J.~P.}\ \bibnamefont {Garrahan}},\
  }\href {http://stacks.iop.org/1742-5468/2014/i=9/a=P09017} {\bibfield
  {journal} {\bibinfo  {journal} {Journal of Statistical Mechanics: Theory and
  Experiment}\ }\textbf {\bibinfo {volume} {2014}},\ \bibinfo {pages} {P09017}
  (\bibinfo {year} {2014})}\BibitemShut {NoStop}%
\bibitem [{\citenamefont {Cammarota}\ and\ \citenamefont
  {Biroli}(2012)}]{Cammarota2012}%
  \BibitemOpen
  \bibfield  {author} {\bibinfo {author} {\bibfnamefont {C.}~\bibnamefont
  {Cammarota}}\ and\ \bibinfo {author} {\bibfnamefont {G.}~\bibnamefont
  {Biroli}},\ }\href@noop {} {\bibfield  {journal} {\bibinfo  {journal} {Proc.
  Natl. Acad. Sci. USA}\ }\textbf {\bibinfo {volume} {109}},\ \bibinfo {pages}
  {8850} (\bibinfo {year} {2012})}\BibitemShut {NoStop}%
\bibitem [{\citenamefont {Jack}\ and\ \citenamefont
  {Berthier}(2012)}]{Jack2012}%
  \BibitemOpen
  \bibfield  {author} {\bibinfo {author} {\bibfnamefont {R.~L.}\ \bibnamefont
  {Jack}}\ and\ \bibinfo {author} {\bibfnamefont {L.}~\bibnamefont
  {Berthier}},\ }\href@noop {} {\bibfield  {journal} {\bibinfo  {journal}
  {Phys. Rev. E}\ }\textbf {\bibinfo {volume} {85}},\ \bibinfo {pages} {021120}
  (\bibinfo {year} {2012})}\BibitemShut {NoStop}%
\bibitem [{\citenamefont {Kob}\ and\ \citenamefont {Berthier}(2013)}]{Kob2013}%
  \BibitemOpen
  \bibfield  {author} {\bibinfo {author} {\bibfnamefont {W.}~\bibnamefont
  {Kob}}\ and\ \bibinfo {author} {\bibfnamefont {L.}~\bibnamefont {Berthier}},\
  }\href@noop {} {\bibfield  {journal} {\bibinfo  {journal} {Phys. Rev. Lett.}\
  }\textbf {\bibinfo {volume} {110}},\ \bibinfo {pages} {245702} (\bibinfo
  {year} {2013})}\BibitemShut {NoStop}%
\bibitem [{\citenamefont {Aizenman}\ and\ \citenamefont
  {Wehr}(1989)}]{Aizenman1989}%
  \BibitemOpen
  \bibfield  {author} {\bibinfo {author} {\bibfnamefont {M.}~\bibnamefont
  {Aizenman}}\ and\ \bibinfo {author} {\bibfnamefont {J.}~\bibnamefont
  {Wehr}},\ }\href@noop {} {\bibfield  {journal} {\bibinfo  {journal} {Phys.
  Rev. Lett.}\ }\textbf {\bibinfo {volume} {62}},\ \bibinfo {pages} {2503}
  (\bibinfo {year} {1989})}\BibitemShut {NoStop}%
\bibitem [{\citenamefont {Sollich}\ and\ \citenamefont
  {Evans}(1999)}]{Sollich1999}%
  \BibitemOpen
  \bibfield  {author} {\bibinfo {author} {\bibfnamefont {P.}~\bibnamefont
  {Sollich}}\ and\ \bibinfo {author} {\bibfnamefont {M.~R.}\ \bibnamefont
  {Evans}},\ }\href@noop {} {\bibfield  {journal} {\bibinfo  {journal} {Phys.
  Rev. Lett.}\ }\textbf {\bibinfo {volume} {83}},\ \bibinfo {pages} {3238}
  (\bibinfo {year} {1999})}\BibitemShut {NoStop}%
\bibitem [{\citenamefont {Jack}\ \emph {et~al.}(2011)\citenamefont {Jack},
  \citenamefont {Hedges}, \citenamefont {Garrahan},\ and\ \citenamefont
  {Chandler}}]{Jack2011}%
  \BibitemOpen
  \bibfield  {author} {\bibinfo {author} {\bibfnamefont {R.~L.}\ \bibnamefont
  {Jack}}, \bibinfo {author} {\bibfnamefont {L.~O.}\ \bibnamefont {Hedges}},
  \bibinfo {author} {\bibfnamefont {J.~P.}\ \bibnamefont {Garrahan}}, \ and\
  \bibinfo {author} {\bibfnamefont {D.}~\bibnamefont {Chandler}},\ }\href@noop
  {} {\bibfield  {journal} {\bibinfo  {journal} {Phys. Rev. Lett.}\ }\textbf
  {\bibinfo {volume} {107}},\ \bibinfo {pages} {275702} (\bibinfo {year}
  {2011})}\BibitemShut {NoStop}%
\bibitem [{\citenamefont {Jack}(2013)}]{Jack2013}%
  \BibitemOpen
  \bibfield  {author} {\bibinfo {author} {\bibfnamefont {R.~L.}\ \bibnamefont
  {Jack}},\ }\href@noop {} {\bibfield  {journal} {\bibinfo  {journal} {Phys.
  Rev. E}\ }\textbf {\bibinfo {volume} {88}},\ \bibinfo {pages} {062113}
  (\bibinfo {year} {2013})}\BibitemShut {NoStop}%
\bibitem [{\citenamefont {Monasson}(1995)}]{Monasson1995}%
  \BibitemOpen
  \bibfield  {author} {\bibinfo {author} {\bibfnamefont {R.}~\bibnamefont
  {Monasson}},\ }\href@noop {} {\bibfield  {journal} {\bibinfo  {journal}
  {Phys. Rev. Lett.}\ }\textbf {\bibinfo {volume} {75}},\ \bibinfo {pages}
  {2847} (\bibinfo {year} {1995})}\BibitemShut {NoStop}%
\bibitem [{\citenamefont {Berthier}\ and\ \citenamefont
  {Jack}()}]{Berthier2015}%
  \BibitemOpen
  \bibfield  {author} {\bibinfo {author} {\bibfnamefont {L.}~\bibnamefont
  {Berthier}}\ and\ \bibinfo {author} {\bibfnamefont {R.~L.}\ \bibnamefont
  {Jack}},\ }\href@noop {} {\bibinfo  {journal} {arXiv:1503.08576}\
  }\BibitemShut {NoStop}%
\bibitem [{\citenamefont {Deng}\ \emph {et~al.}(2010)\citenamefont {Deng},
  \citenamefont {Guo}, \citenamefont {Heringa}, \citenamefont {Bl{\"o}te},\
  and\ \citenamefont {Nienhuis}}]{Deng2010}%
  \BibitemOpen
\bibfield  {journal} {  }\bibfield  {author} {\bibinfo {author} {\bibfnamefont
  {Y.}~\bibnamefont {Deng}}, \bibinfo {author} {\bibfnamefont {W.}~\bibnamefont
  {Guo}}, \bibinfo {author} {\bibfnamefont {J.~R.}\ \bibnamefont {Heringa}},
  \bibinfo {author} {\bibfnamefont {H.~W.}\ \bibnamefont {Bl{\"o}te}}, \ and\
  \bibinfo {author} {\bibfnamefont {B.}~\bibnamefont {Nienhuis}},\ }\href@noop
  {} {\bibfield  {journal} {\bibinfo  {journal} {Nucl. Phys. B}\ }\textbf
  {\bibinfo {volume} {827}},\ \bibinfo {pages} {406} (\bibinfo {year}
  {2010})}\BibitemShut {NoStop}%
\bibitem [{\citenamefont {Xu}\ and\ \citenamefont {Moore}(2004)}]{Xu2004}%
  \BibitemOpen
  \bibfield  {author} {\bibinfo {author} {\bibfnamefont {C.}~\bibnamefont
  {Xu}}\ and\ \bibinfo {author} {\bibfnamefont {J.~E.}\ \bibnamefont {Moore}},\
  }\href@noop {} {\bibfield  {journal} {\bibinfo  {journal} {Phys. Rev. Lett.}\
  }\textbf {\bibinfo {volume} {93}},\ \bibinfo {pages} {047003} (\bibinfo
  {year} {2004})}\BibitemShut {NoStop}%
\bibitem [{\citenamefont {Xu}\ and\ \citenamefont {Moore}(2005)}]{Xu2005}%
  \BibitemOpen
  \bibfield  {author} {\bibinfo {author} {\bibfnamefont {C.}~\bibnamefont
  {Xu}}\ and\ \bibinfo {author} {\bibfnamefont {J.~E.}\ \bibnamefont {Moore}},\
  }\href@noop {} {\bibfield  {journal} {\bibinfo  {journal} {Nucl. Phys. B}\
  }\textbf {\bibinfo {volume} {716}},\ \bibinfo {pages} {487 } (\bibinfo {year}
  {2005})}\BibitemShut {NoStop}%
\end{thebibliography}
\end{document}